\documentclass[11point]{article}

\usepackage{color,epsfig,amssymb,amsmath,subfigure,url}
\usepackage{changepage}
\usepackage{bm}
\usepackage{algorithm}
\usepackage{algorithmic}
\usepackage[normalem]{ulem}
\usepackage{enumerate}

\renewcommand{\thefootnote}{\fnsymbol{footnote}}
\title{Detecting edges from non-uniform Fourier data via sparse Bayesian learning}
\author{Victor Churchill \footnotemark[1] \and Anne Gelb \footnotemark[1]}

\begin{document}
\footnotetext[1]{Department of Mathematics, Dartmouth College, USA}

\newtheorem{thm4}{Theorem}
\newtheorem{thm5}{Theorem}
\newtheorem{thm}{Theorem}
\newtheorem{defn}[thm5]{Definition}
\newtheorem{example}[thm4]{Example}
\renewcommand{\thefootnote}{\arabic{footnote}}

\maketitle

\section*{Abstract}

In recent investigations, the problem of detecting edges given non-uniform Fourier data was reformulated as a sparse signal recovery problem with an $\ell_1$-regularized least squares cost function. This result can also be derived by employing a Bayesian formulation.  Specifically, reconstruction of an edge map using $\ell_1$ regularization corresponds to a so-called type-I (maximum \textit{a posteriori}) Bayesian estimate. In this paper, we use the Bayesian framework to design an improved algorithm for detecting edges from non-uniform Fourier data. In particular, we employ what is known as type-II Bayesian estimation, specifically a method called sparse Bayesian learning. We also show that our new edge detection method can be used to improve downstream processes that rely on accurate edge information like image reconstruction, especially with regards to compressed sensing techniques.

\section{Introduction}
\label{sec:intro}

Edge detection is an important tool in identifying physical structures and regions of interest in signals and images. In magnetic resonance imaging (MRI), edge detection helps tissue boundary identification, \cite{jimenez2006data,shattuck2001magnetic}. In synthetic aperture radar (SAR), it can improve target identification. In both of these applications, data are collected as non-uniform Fourier samples. Detecting edges specifically from non-uniform\footnote{Note that while here we only explicitly consider non-uniform Fourier samples, all methods described here apply to uniform Fourier samples as well.} Fourier data has been explored in \cite{gelb2011detection,gelb2017detecting,martinez2014edge}. Most recently, the fact that the edges of many signals and images are sparse was utilized to reformulate edge detection from non-uniform Fourier data as a sparse signal recovery (SSR) problem, \cite{gelb2017detecting}. Groundwork for that reformulation was laid in \cite{gelb2011detection,martinez2014edge,stefan2012sparsity,viswanathan2012iterative}. The technique in \cite{gelb2017detecting} used Fourier frames to construct a forward model for edge detection from given non-uniform Fourier data.  An edge map of a piecewise smooth function was then reconstructed using an $\ell_1$-norm regularization optimization procedure which is known to encourage sparsity, \cite{chen2001atomic,tibshirani1996regression}.\footnote{Although ideally the $\ell_0$ semi-norm should be used to regularize this problem, the resulting optimization problem is NP-hard. Hence the $\ell_1$ norm has become a  popular convex surrogate that makes the problem computationally tractable and also offers theoretical guarantees for exact reconstruction, \cite{candes2006robust}, as well as a variety of other benefits related to compressed sensing, \cite{donoho2006compressed}.} For a variety of reasons including noise, non-uniform sampling, and the magnitude dependence of the $\ell_1$ norm, the solution via $\ell_1$ regularization to SSR is sometimes not as sparse as desired. While the edge reconstructions in \cite{gelb2017detecting} may be suboptimal due to these downsides of $\ell_1$ regularization, it was indeed a key development of \cite{gelb2017detecting}, although not explicitly addressed, to view edge detection as SSR. With that door open, in this paper we explore another algorithm that has both empirically and theoretically outperformed $\ell_1$ regularization for SSR.

Capturing the sparsity of solutions more accurately in the SSR problem has been widely studied, \cite{babacan2010bayesian,candes2006robust,candes2008enhancing,chartrand2008iteratively,giri2016type,ji2008bayesian,tipping2001sparse,wipf2004sparse}. More recently, there has been increased interest in Bayesian probabilistic approaches to SSR, \cite{babacan2010bayesian,giri2016type,ji2008bayesian,tipping2001sparse,wipf2004sparse}. Within the probabilistic approaches there are two categories. The first is type-I, or maximum \textit{a posteriori} (MAP) Bayesian estimation which uses a fixed prior. The most popular examples of type-I methods for SSR are $\ell_1$ regularization, \cite{chen2001atomic,tibshirani1996regression}, iteratively reweighted $\ell_1$ regularization, \cite{candes2008enhancing}, and iteratively reweighted $\ell_2$ regularization, \cite{chartrand2008iteratively}. The second category is type-II, or evidence maximization Bayesian estimation which employs a hierarchical, flexible parametrized prior that is {\em learned} from the given data. For SSR, \cite{giri2016type} provides the most general and comprehensive comparison of type-I and type-II methods,  and in particular it is shown there through extensive empirical results that type-II estimates outperform corresponding type-I estimates in terms of accuracy. In addition, type-II methods have the ability to retrieve a full posterior distribution for the solution rather than just a point estimate as in type-I. Also advantageous is the automatic estimation of the crucial regularization parameter, the choice of which is typically difficult and subjective in MAP schemes. In this paper we focus our efforts on a particular type-II method called sparse Bayesian learning (SBL), also called the relevance vector machine (RVM), \cite{tipping2001sparse}. In \cite{giri2016type}, SBL typically outperformed all other methods including $\ell_1$ regularization, \cite{candes2006robust}, iteratively reweighted $\ell_1$, \cite{candes2008enhancing}, and iteratively reweighted $\ell_2$ regularization methods, \cite{chartrand2008iteratively}, as well as a variety of type-II methods for SSR. The strength of SBL is also supported by evidence from \cite{ji2008bayesian}, which looked at SBL from the perspective of compressed sensing, \cite{candes2006robust,donoho2006compressed}, and showed that SBL could recover a sparse signal with better accuracy than $\ell_1$ regularization using the same amount of data. There are also theoretical results from \cite{rao2006comparing,wipf2005norm} that show that SBL is a closer surrogate to the $\ell_0$ norm than $\ell_1$ regularization, and proved that even in the worst case SBL still outperforms the most widely employed algorithms from compressed sensing. In \cite{wipf2004sparse}, for the noiseless case, it is proved that the global minimum of the SBL cost function is achieved at a solution such that the posterior mean equals the maximally sparse solution. Furthermore, it was shown that local minima are achieved at sparse solutions, regardless of noise, \cite{wipf2004sparse}. SBL employs the expectation-maximization (EM) algorithm, \cite{dempster1977maximum}. Due to properties of the EM algorithm, SBL is globally convergent, \cite{wipf2004sparse}.

This paper adapts and implements the SBL algorithm from \cite{tipping2001sparse} for the non-uniform Fourier edge detection problem using the SSR formulation from \cite{gelb2017detecting}. We demonstrate improved accuracy compared with the results of \cite{gelb2017detecting} that used $\ell_1$ regularization in both noise-free and noisy scenarios. The organization of the paper is as follows. In Section \ref{sec:setup}, we set up the problem of edge detection from non-uniform Fourier data. In Section \ref{sec:formulating}, we describe the reformulation of the problem as SSR and reproduce the edge detection results from \cite{gelb2017detecting}. Section \ref{probabilistic} introduces the probabilistic approach, adapts SBL for the edge detection problem, and explains the resulting algorithm. We also look at numerical results for the new SBL technique, comparing with the results of Section \ref{sec:formulating}. Edge detection is not only useful in and of itself, but can be used to improve signal and image reconstruction as well. In Section \ref{sec:applications}, we use our new edge detection algorithm to create a weighting matrix in order to apply $\ell_2$ regularization away from edges to achieve full image reconstruction results. This is so-called edge-adaptive $\ell_2$ regularization, \cite{churchill2018edge}. Finally in Section \ref{sec:conclusion}, we summarize this work and look at future directions for our research.

\section{Problem Setup}
\label{sec:setup}

We consider a one-dimensional piecewise smooth function $f:[-1,1]\rightarrow\mathbb{R}$. We define the jump function, $[f]$, of $f$ as the difference between the left- and right-hand limits of the function:
\begin{align}\label{jumpdef}
[f](x) &= f(x^+)-f(x^-).
\end{align}
In smooth regions, $[f](x)=0$. At a discontinuity, $[f](x)$ is equal to the height of the jump. Suppose we use $2J+1$ grid points, $x_j = \frac{j}{J}$ for $j=-J,\ldots,J$. Assuming that the discontinuities of $f$ are separated such that there is at most one jump per cell $I_j=[x_j,x_{j+1})$, we can write
\begin{align}
\label{eq:jumpdelta}
[f](x) & = \sum_{j=-J}^{J-1}[f](x_j)\delta_{x_j}(x).
\end{align}
where the coefficient $[f](x_j)$ is the jump value in $I_j$ and $\delta_{x_j}(x)$ is the indicator function with 
\begin{align}
\delta_{x_j}(x) =  \left\{
\begin{array}{lr}
1 & x = x_j\\
0 & x\neq x_j
\end{array}
\right..
\end{align}

\begin{figure}[h!]
\centering
\includegraphics[width=\textwidth]{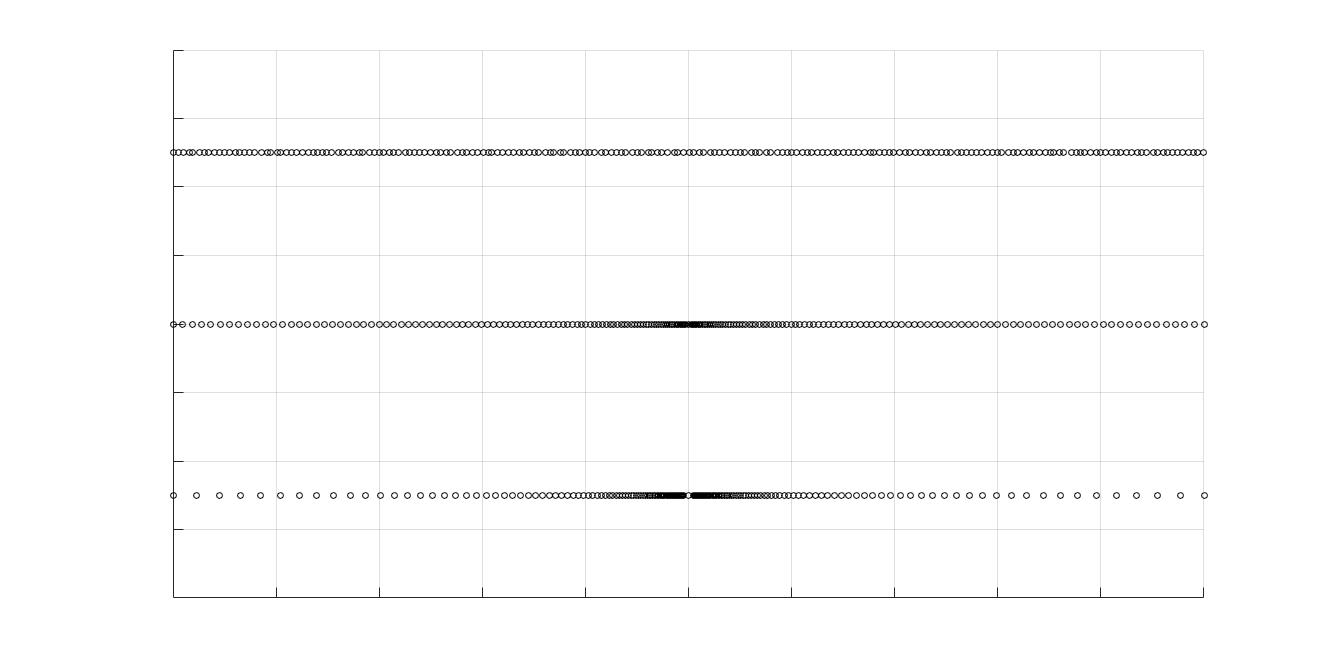} 
\caption{Modes $\mathbf{\lambda}_{k}$ given by (top) jittered; (middle) quadratic; and (bottom) logarithmic sampling.}
\label{fig:1D_sampling}
\end{figure}

Suppose we are given a finite sequence of non-uniform Fourier samples of $f$,
\begin{eqnarray}
\label{eq:fourierdata}
\hat{f}(\lambda_k) & = & \frac{1}{2}\int_{-1}^1f(x)e^{-i\pi\lambda_kx}dx,
\end{eqnarray}
where $\lambda_k\in\mathbb{R}$ and $k=-M,\ldots,M$. Note here that we are considering \textit{continuous} Fourier samples, as generating data via the same model we use to solve the inverse problem commits the inverse crime. Specifically, we look at three types of non-uniform sampling that were considered in \cite{gelb2017detecting}. The first is \textit{jittered sampling}, defined by
\begin{eqnarray}
\label{eq:jittered}
\lambda_k & = & k-\left\lfloor \frac{2M+1}{2} \right\rfloor -1+\frac{1-2\xi_k}{4}
\end{eqnarray}
where $\xi_k \sim U([0,1])$. Jittered sampling imitates Cartesian grid sampling with small errors that often occur in imaging systems. The next is \textit{quadratic sampling}, defined by
\begin{align}\label{eq:quadratic}
\lambda_k &= \frac{\text{sign}(k)k^2}{M}.
\end{align}
This pattern simulates a cross-section of a non-Cartesian sampling pattern that undersamples high frequencies and oversamples low frequencies. Lastly, we consider \textit{logarithmic sampling}, which oversamples low frequencies and, even more sparsely than quadratic, undersamples high frequencies. In particular, $\log|\lambda_k|$ is evenly distributed between $-v$ and $\log n$ with $v>0$ and $2n+1$ being the total number of samples. Figure \ref{fig:1D_sampling} gives a visualization of these three sampling types. We will also consider the case where the underlying Fourier data in (\ref{eq:fourierdata}) are noisy, given by
\begin{equation}
\label{eq:noisydata}
\hat{f}^\eta (\lambda_k) = \hat{f}(\lambda_k) + \eta_k,
\end{equation}
for $k = -M,\cdots,M$.  Here $\eta_k\sim\mathcal{CN}(0,\sigma^2)$, meaning $\eta_k$ is a complex Gaussian random variable with mean $0$ and variance $\sigma^2$.

From data given by (\ref{eq:fourierdata}) or (\ref{eq:noisydata}), our goal is to accurately reconstruct the edge function $[f](x)$ defined in \eqref{jumpdef}. In the following section, we review the method of \cite{gelb2017detecting}, which shows how non-uniform Fourier edge detection can be formulated as SSR. Later, we see the same formulation in probablistic terms, and then go further into the SBL framework to achieve superior results.

\section{Formulating edge detection as SSR}
\label{sec:formulating}

In \cite{gelb2017detecting}, the authors rely on an optimization to choose a concentration factor, \cite{gelb1999detection}, followed by a reconstruction via optimizing an $\ell_1$-regularized cost function. The $\ell_1$ regularized edge detection method described below closely follows the derivation in \cite{gelb2017detecting}.

\subsection{Concentration factor edge detection for uniform Fourier data}
The concentration factor edge detection method, originally developed in \cite{gelb1999detection}, approximates (\ref{eq:jumpdelta}) from $2M+1$ uniform Fourier coefficients given in (\ref{eq:fourierdata}) where $\lambda_k=k$ as
\begin{align}
\label{eq:cf}
S_M^\sigma[f](x)& = i\sum_{k=-M}^{M}\hat{f}(k)\text{sgn}(k)\sigma\left(\frac{|k|}{M}\right) e^{\pi ikx}.
\end{align}
Here ${\bf \sigma}$, coined the concentration factor in \cite{gelb1999detection}, satisfies certain admissibility conditions
\begin{enumerate}
\item $\sum_{k=1}^M\sigma\left(\frac{|k|}{M}\right)\sin(k\pi x)$ is odd
\item $\frac{\sigma(u)}{u}\in C^2(0,1)$
\item $\int_\nu^1\frac{\sigma(u)}{u}du\rightarrow-1$, $\nu=\nu(M)>0$ being small.
\end{enumerate}
If all three conditions are satisfied, $S_M^\sigma [f]$ concentrates at the singular support of $f$ and the jump function approximation observes the concentration property
\begin{eqnarray}
\label{eq:concprop}
S_M^\sigma[f](x)& = & [f](x) + \left\{
\begin{array}{ll}
      \mathcal{O}\left(\frac{\log M}{M}\right) & d(x) \le \frac{\log M}{M}\\
      \mathcal{O}\left(\frac{\log M}{(Md(x))^s}\right) & d(x) >>\frac1M
\end{array} 
\right.
\end{eqnarray}
where $d(x)$ is the distance between a point in the domain and the nearest discontinuity, and $s>0$ depends on $\sigma$. Generally, the convergence of (\ref{eq:cf}) depends on the particular choice of ${\bf \sigma}$. 

\subsection{Fourier Frame Approximation}
The concentration factor method cannot be extended directly to non-uniform Fourier coefficients because $\{e^{i\pi\lambda_k x}\}_{k=-M}^M$ is generally not an orthogonal basis.  However, it can be adapted by using the finite Fourier frame approximation, \cite{gelb2017detecting,song2013approximating}. We will employ the Fourier frame framework for our new edge detection method.  The technique in \cite{gelb2017detecting} is briefly described below.

We require the following two definitions:
\begin{defn}
A \textbf{frame} for a Hilbert space $\mathcal{H}$ is a sequence of vectors $\{\varphi_k:k\in\mathbb{Z}\}\subseteq \mathcal{H}$ for which there exists constants $0<A\le B<\infty$ such that, for every $f\in \mathcal{H}$, we have
\begin{align}\label{eq:frame}
A||f||^2\le \sum_{k\in\mathbb{Z}}|\langle f,\varphi_k\rangle|^2\le B||f||^2.
\end{align}
\end{defn}

\begin{defn}
If $\{\varphi_k:k\in\mathbb{Z}\}\subseteq\mathcal{H}$ is a frame for $\mathcal{H}$ then the associated \textbf{frame operator} $S:\mathcal{H}\rightarrow\mathcal{H}$ is defined as
\begin{align}\label{eq:frameoperator}
Sf =\sum_{k\in\mathbb{Z}}\langle f,\varphi_k\rangle\varphi_k.
\end{align}
\end{defn}

Frame elements span $\mathcal{H}$ but are not necessarily linearly independent. The frame operator is bounded, invertible, positive, and self-adjoint. In this investigation we consider $\mathcal{H}=L^2(-1,1)$ and $\varphi_k(x) = e^{i\pi\lambda_k x}$. Hence we can reconstruct $f$ via
\begin{align}\label{eq:framerecon}
f = S^{-1}Sf = \sum_{k\in\mathbb{Z}} \langle f,\varphi_k\rangle S^{-1}\varphi_k = \sum_{k\in\mathbb{Z}} \hat{f}(\lambda_k)\tilde{\varphi}_k,
\end{align}
where $\tilde{\varphi}_k=S^{-1}\varphi_k$ for $k\in\mathbb{Z}$ is the canonical dual frame and $\hat{f}(\lambda_k)$ is the given non-uniform Fourier data as in (\ref{eq:fourierdata}). In general there is no closed form of $S^{-1}$. Various algorithms construct (finite-dimensional) approximations to $S^{-1}$. Here we use the {\em admissible frame} method developed in \cite{song2013approximating}, which obtains a finite-dimensional approximation of $\tilde{\varphi}_k$ by projecting $\{\varphi_k:k\in\mathbb{Z}\}$ onto an admissible frame $\{\psi_l:l\in\mathbb{Z}\}$ defined as

\begin{defn}
A frame $\{\psi_l:l\in\mathbb{Z}\}$ is \textbf{admissible} with respect to the frame $\{\varphi_k:k\in\mathbb{Z}\}$ if the following two conditions hold:
\begin{enumerate}
\item It is intrinsically self-localized, that is, there exists $c_0\in\mathbb{R}^+$ and $t>1$ such that $$|\langle\psi_k,\psi_l\rangle|\le c_0(1+|k-l|)^{-t}.$$
\item There exists $c_1\in\mathbb{R}^+$ and $s>1/2$ such that $$|\langle \varphi_k,\psi_l\rangle|\le c_1(1+|k-l|)^{-s}.$$
\end{enumerate}
\end{defn}

If $\{\psi_l:l\in\mathbb{Z}\}$ is admissible with respect to $\{\varphi_k:k\in\mathbb{Z}\}$, then $\{\tilde{\varphi}_k:k\in\mathbb{Z}\}$ can be approximated by
\begin{align}\label{eq:dualframeapprox}
\tilde{\varphi}_k\approx\sum_{l=-N}^N b_{l,k}\psi_l =:\tilde{\varphi}_{N,k},
\end{align}
where $\{b_{l,k}\}_{l=-N,k=-M}^{N,M}$ are the entires of $\mathbf{B}=\Psi^\dagger$. That is, $\mathbf{B}$ is the Moore-Penrose psuedoinverse of $\Psi$ where
\begin{align}\label{eq:Psi}
\Psi_{k,l} = [\langle\varphi_k,\psi_l\rangle]\quad k=-M,\ldots,M,\quad l=-N,\ldots,N.
\end{align}
Since in this investigation $\varphi_k(x) = e^{i\pi\lambda_kx}$, a practical admissible frame is $\psi_l(x)= e^{i\pi lx}$, yielding the $sinc$ approximation
\begin{align}
\langle\varphi_k,\psi_l\rangle = 2\frac{\sin(\pi(\lambda_k-l))}{\pi(\lambda_k-l)}.
\end{align}
Since the conditioning of (\ref{eq:Psi}) depends on the difference $\lambda_k - l$, more uniformly spaced samples yield better conditioning. To complete the reconstruction of $f$, we plug in the approximation of $\tilde{\varphi}_k$ in (\ref{eq:dualframeapprox}) into (\ref{eq:framerecon}) to get
\begin{align}\label{eq:frameapproxrecon}
T_Mf = \sum_{l=-N}^N\sum_{k=-M}^M\hat{f}(\lambda_k)b_{l,k}\psi_l=\sum_{k=-M}^M\hat{f}(\lambda_k)\tilde{\varphi}_{N,k}(x).
\end{align}

\subsection{Concentration factor design}

Returning to detecting edges from non-uniform Fourier data, since we cannot use (\ref{eq:cf}), we instead use (\ref{eq:frameapproxrecon}) to approximate (\ref{eq:jumpdelta}) by
\begin{align}\label{eq:Tsigma}
T_M^\sigma f(x) = \sum_{k=-M}^M\sigma_k\hat{f}(\lambda_k)\tilde{\varphi}_{N,k}(x).
\end{align}
The problem of choosing a concentration factor still remains. We need $\sigma\in\mathbb{C}^{2M+1}$ so that for each reconstruction point $x_j=\frac jJ$, $j=-J,\ldots,J$, we have
\begin{align}\label{eq:Tsigmaapprox}
T_M^\sigma f(x_j) = \sum_{k=-M}^M \sigma_k\hat{f}(\lambda_k)\tilde{\varphi}_{N,k}(x_j)\approx[f](\xi)\delta_{\xi}(x_j).
\end{align}
An optimization problem for determining $\sigma$ is developed in \cite{gelb2017detecting}.   This is accomplished by first approximating $f$ as a superposition of scaled and shifted ramp functions.  Specifically, define the ramp function $r:[-1,1]\rightarrow\mathbb{R}$ as
\begin{align}\label{eq:ramp}
r(x) = \left\{\begin{array}{cc}
	-\frac{x+1}{2} &\mbox{ if $x\le0$}\\
	-\frac{x-1}{2} &\mbox{ if $x>0$}
\end{array}
\right.,
\end{align}
and observe that the jump function for $r_\xi(x) = r(x-\xi)$ for $\xi\in(-1,1)$ is
\begin{align}\label{eq:rampjump}
[r_\xi](x) = \left\{\begin{array}{cc}
	1 &\mbox{ if $x=\xi$}\\
	0 & \mbox{ if $x\neq\xi$}
\end{array}
\right..
\end{align}
A first order approximation for $f(x)$ with a single jump at $x=\xi$ is then
\begin{align}
f(x)\approx ar_\xi(x)
\end{align}
where $a$ is the jump height. For multiple jumps, we use (\ref{eq:ramp}) and (\ref{eq:rampjump}) to obtain
\begin{align}\label{eq:firstorderapprox}
f(x)\approx\sum_{j=-J}^{J-1}a_jr_{\xi_j}(x)
\end{align}
where $a_j$ is the height of the jump in cell $I_j$. 

To determine the concentration vector $\{\sigma_k\}_{k = -M}^M$,  we can consider the case when there is a single jump discontinuity at $x = \xi$.\footnote{Due to linearity of (\ref{eq:Tsigmaapprox}) on the Fourier data, the following results for functions with a single jump will also hold for a function with multiple jumps.}
First, we discretize $ar_\xi(x)$, substitute into (\ref{eq:Tsigmaapprox}) and translate to get
\begin{align}\label{eq:deltaapprox}
\sum_{k=-M}^M \sigma_k\hat{r}(\lambda_k)\tilde{\varphi}_{N,k}(x_j)\approx \delta_0(x_j).
\end{align}
Here $\hat{r}(\lambda_k)$ are the Fourier coefficients for $r(x)$ given by
\begin{align}
\hat{r}(\lambda_k) = \left\{\begin{array}{cc}
	0 & \lambda_k=0\\
	\frac{(\sin(\pi\lambda_k)-\pi\lambda_k)i}{(\pi\lambda_k)^2} & \lambda_k\neq0.
\end{array}
\right.
\end{align}
Because $\delta_\xi(x)$ has a trivial Fourier expansion, we consider a regularized approximation of $\delta_\xi(x)$,
\begin{align}
h_\xi(x) &=\delta_\xi(x),\quad \epsilon>0,
\end{align}
where $h_\xi(x) = h(\frac{x-\xi}{\epsilon})$ for some $h$ that is (essentially) compactly supported in $[\xi-\epsilon,\xi+\epsilon]$ with $h_\xi(\xi)=1$. As the parameter $\epsilon$ increases the approximation is more regularized, but the edges are not as well localized. The authors of \cite{gelb2017detecting} suggest using $\epsilon=.07$. Replacing $\delta_0(x)$ in (\ref{eq:deltaapprox}) gives
\begin{align}
\sum_{k=-M}^M\sigma_k\hat{r}(\lambda_k)\tilde{\varphi}_{N,k}(x_j)\approx h_0(x_j) \approx \sum_{k=-M}^M\hat{h}_0(\lambda_k)\tilde{\varphi}_{N,k}(x_j).
\end{align}
Hence
\begin{align}
\sigma_k = \frac{\hat{h}_0(\lambda_k)}{\hat{r}(\lambda_k)},\quad k=-M,\ldots,M.
\end{align}
Now we know how to compute $\sigma$, but the problem of choosing an optimal $h$ still remains. In many cases, $f$ is in a more restrictive class of functions like piecewise polynomials or piecewise trigonometric polynomials. Assuming $f$ has just one edge at $x=\xi$,
\begin{align}
f(x) = s(x) + [f](\xi)r_\xi(x)
\end{align}
where $s(x)$ is the continuous part of $f$. By adding the constraint 
$$T_M^\sigma s(x_j)\approx0$$ 
for all $x_j\in(-1,1)$, we are requiring that smooth regions of $f$ be drawn to zero in the approximation of $[f](x)$. With these tools in hand, we can now develop an algorithm to construct an optimal $\mathbf{\hat{h}}=\{\hat{h}_0(\lambda_k)\}_{k=-M}^M$ by formulating an optimization problem rather than using a pre-determined regularization that makes no assumption on $s(x)$. Specifically, we determine that $\mathbf{\hat{h}}$ to satisfy two contraints 
\begin{align}\label{eq:constraints}
\sum_{k=-M}^M\frac{\hat{h}(\lambda_k)}{\hat{r}(\lambda_k)}\hat{s}(\lambda_k)\tilde{\varphi}_{N,k}(x_j)\approx0\quad\text{and}\quad\sum_{k=-M}^M\hat{h}(\lambda_k)\tilde{\varphi}_{N,k}(x_j)\approx\delta_0(x_j).
\end{align}

\begin{algorithm}[htbp!]
\caption{Concentration factor design for non-uniform Fourier data}
\label{alg:cfdesign}
\begin{algorithmic}[1]
\STATE Given $2M+1$ Fourier coefficients, $\hat{f}(\lambda_k)$, of a piecewise smooth function, $f$, as in (\ref{eq:fourierdata}).

\STATE Choose $s(x)$ to be consistent with $f(x) = s(x) + [f](\xi)r_\xi(x)$ and define
\begin{align}\label{FS}
\mathbf{F} &= \left[\tilde{\varphi}_{N,k}(x_j)\right]_{j=-J,k=-M}^{J,M}\quad\text{and}\quad \mathbf{S} = \left[\frac{\hat{s}(\lambda_k)}{\hat{r}(\lambda_k)} \tilde{\varphi}_{N,k}(x_j)\right]_{j=-J,k=-M}^{J,M}.
\end{align}

\STATE Determine $\mathbf{\hat{h}}$ as the minimizer of
\begin{align}\label{cfopt}
\min_{\mathbf{\hat{h}}\in\mathbb{R}^{2M+1}} ||\mathbf{F}\mathbf{\hat{h}}||_1+\mu||\mathbf{S}\mathbf{\hat{h}}||_1
\end{align}
for $x_j=\frac jJ$, $j=-J,\ldots,J$, and $\mu>0$.

\STATE Define $\sigma_k=\frac{\hat{h}_k}{\hat{r}(\lambda_k)}$, $k=-M,\ldots,M$.
\end{algorithmic}
\end{algorithm}
Per \cite{gelb2017detecting}, we choose $\mu=1000$ for Algorithm \ref{alg:cfdesign}. We still have to choose $s(x)$, though, which depends on prior information about $f$. For example, if $f$ has discontinuities in its derivative as well, we may choose a hat function
\begin{align}\label{eq:hat}
s_1(x) = \left\{\begin{array}{cc}
-\frac{x+1}{2} & \mbox{ if $-1\le x\le0$}\\
\frac{x-1}{2} & \mbox{ if $0<x\le1$}
\end{array}\right..
\end{align}
On the other hand if $f$ is essentially piecewise constant or linear with smooth variation between jumps, a smooth varying function like
\begin{align}\label{eq:s_2}
s_2(x) = \left\{\begin{array}{cc}
-\frac{(x+1)^3}{12} & \mbox{ if $-1\le x\le0$}\\
\frac{(x-1)^3}{12}-\frac16 & \mbox{ if $0<x\le1$}
\end{array}\right.,
\end{align}
is more appropriate.

\subsection{Sparsity model for edge detection}\label{sparsitymodel}

We now  develop a method for reconstructing the edge function $[f](x)$ using Algorithm \ref{alg:cfdesign} that takes advantage of our assumption that $[f](x)$ is sparse. We seek a solution $[f](x)$ that most closely matches the projection of $\sum_{l=1}^L[f](\xi_l)h_{\xi_l}(x)$ where $L$ is the number of edges onto the space spanned by the Fourier frame elements given by (\ref{eq:Tsigmaapprox}). To this end, we adopt the waveform matching idea developed in  \cite{stefan2012sparsity} for uniform Fourier data.  As will be evident in what follows, this approach allows us to build a forward model for reconstructing $[f](x)$. We start by defining the waveform kernel as the partial Fourier sum approximation of $h_0(x)$ given by
\begin{align}
W_M^\sigma(x) = \frac{1}{\gamma_M^\sigma}\sum_{k=-M}^M\sigma_k\frac{\cos kx}{k},
\end{align}
where $\gamma_M^\sigma$ is a normalization constant.   We can then approximate
\begin{align}\label{eq:waveform}
W_M^\sigma\ast[f]\approx S_M^\sigma(f),
\end{align}
where $S_M^\sigma(f)$ is defined in (\ref{eq:cf}) for some admissible $\sigma$. For non-uniform Fourier data, the corresponding waveform kernel is given by
\begin{align}
W_{N(M)}^\sigma(x) = \frac{1}{\gamma_{N(M)}^\sigma}\sum_{k=-M}^M\sigma_k\hat{r}(\lambda_k)\tilde{\varphi}_{N,k}(x)= \frac{1}{\gamma_{N(M)}^\sigma}\sum_{k=-M}^M\sum_{l=-N}^N\sigma_k\hat{r}(\lambda_k)b_{l,k}e^{i\pi lx},
\end{align}
where $$\gamma_{N(M)}^\sigma = \sum_{k=-M}^M\sum_{l=-N}^N b_{l,k}\sigma_k\hat{r}(\lambda_k)$$ is the normalization constant. Analogous to (\ref{eq:waveform}) we now have
\begin{align}
W_{N(M)}^\sigma\ast[f]\approx T_M^\sigma(f),
\end{align}
which is satisfied by requiring
\begin{align}
\frac{1}{\gamma_{N(M)}^\sigma}\sum_{k=-M}^M b_{l,k}\sigma_k\hat{r}(\lambda_k)\widehat{[f]}(l)\approx\sum_{k=-M}^Mb_{l,k}\sigma_k\hat{f}(\lambda_k)\quad l=-N,\ldots,N.
\end{align}
Hence we see that using the waveform kernel approach allows us to construct the model
\begin{align}
\frac{1}{\gamma_{N(M)}^\sigma} (\mathbf{B}(\sigma\cdot\mathbf{\hat{r}}))\cdot (\mathbf{F}[f]) \approx \mathbf{B}(\sigma\cdot\mathbf{\hat{f}})
\end{align}
where $\cdot$ denotes elementwise multiplication and $\mathbf{F}$ is the Fourier transform. We simplify the model as
\begin{align}\label{eq:model}
\Theta[f] \approx \mathbf{y}
\end{align}
where
\begin{align}
\Theta = \frac{1}{\gamma_{N(M)}^\sigma}\text{diag}\left[\mathbf{B}(\sigma\cdot\mathbf{\hat{r}})\right] \mathbf{F} \quad \text{and} \quad\mathbf{y} = \mathbf{B}(\sigma\cdot\mathbf{\hat{f}}).
\end{align}

\begin{figure}[h!]
\centering
\includegraphics[width=.32\textwidth]{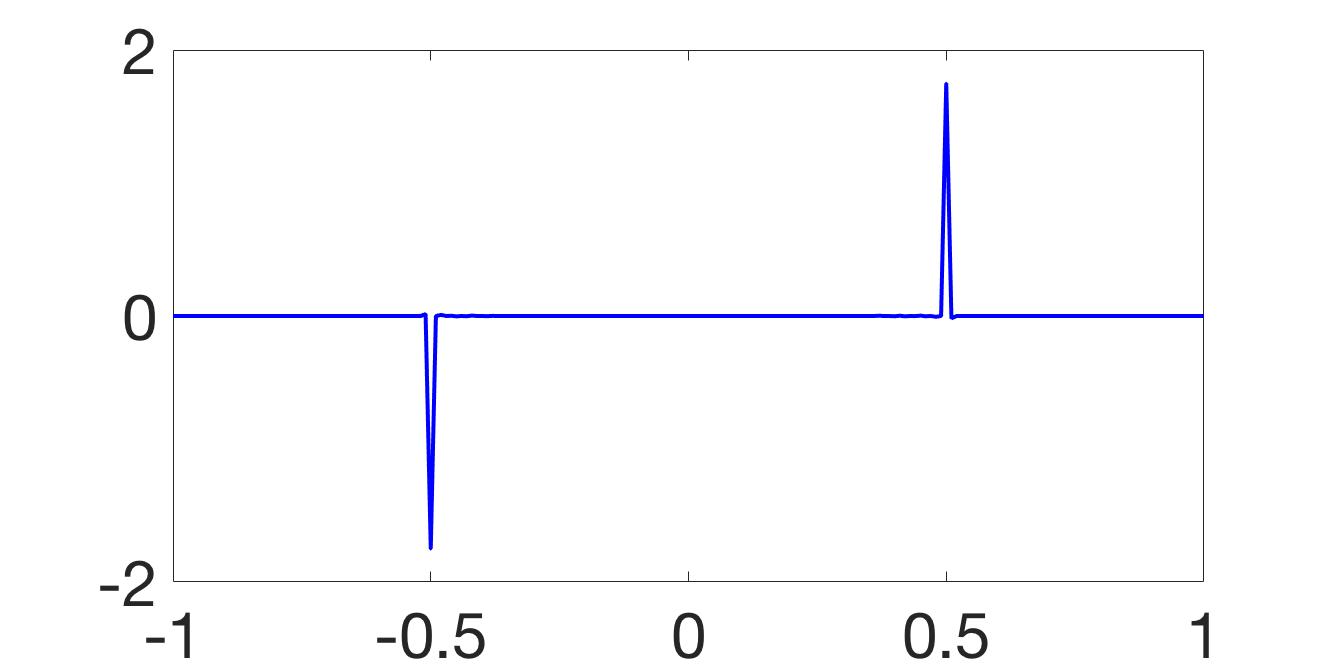}
\includegraphics[width=.32\textwidth]{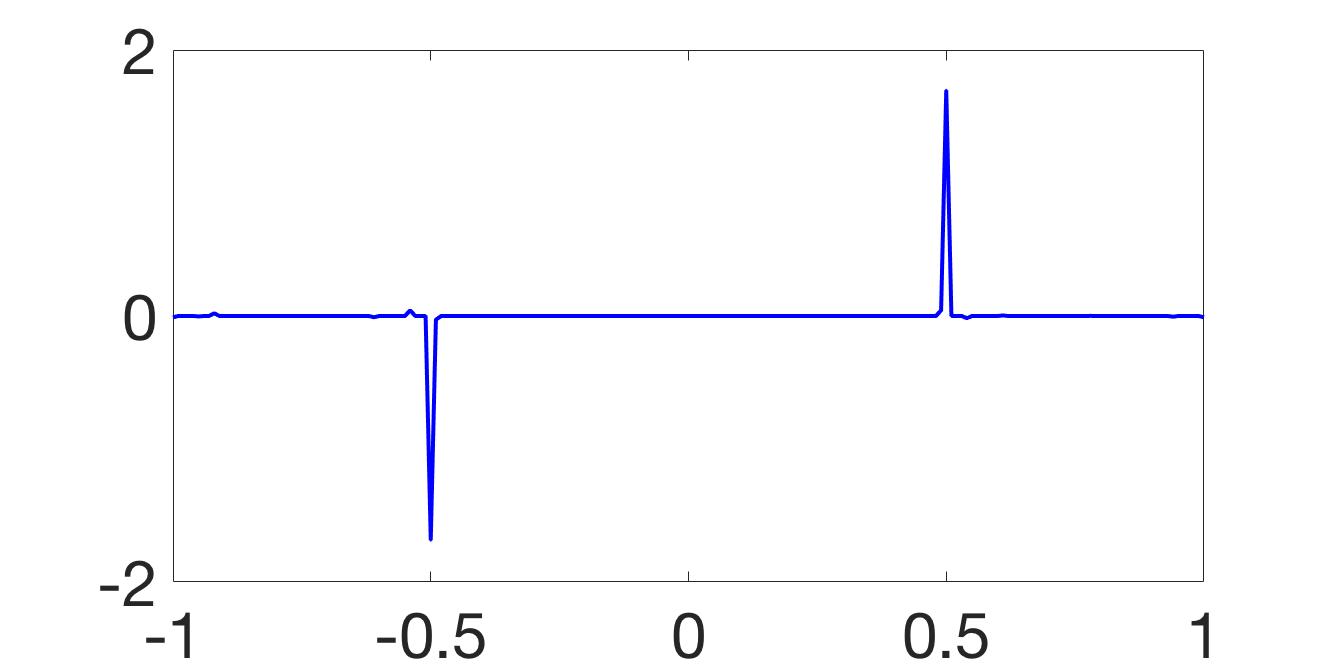}
\includegraphics[width=.32\textwidth]{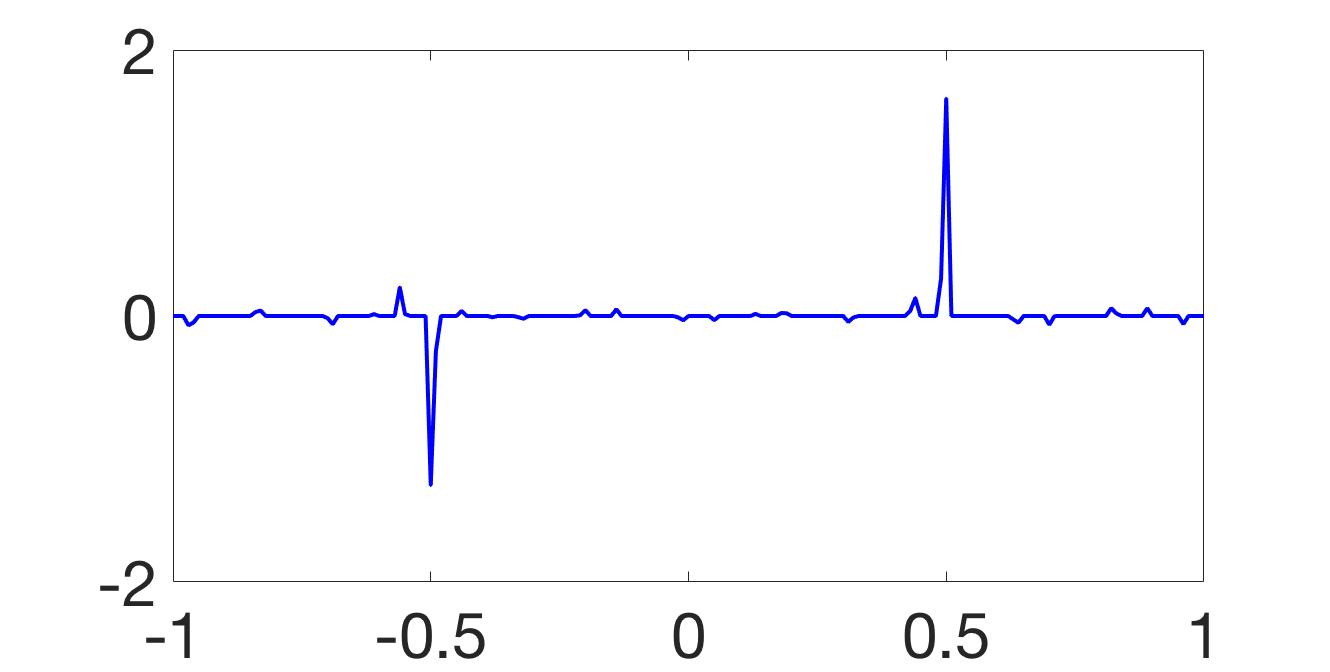}
\caption{$[f_3]$ reconstructed via Algorithm \ref{alg:edgel1} using (left) jittered; (center) quadratic; and (right) logarithmic sampling.}
\label{fig:f3_l1}
\end{figure}

\begin{figure}[h!]
\centering
\includegraphics[width=.32\textwidth]{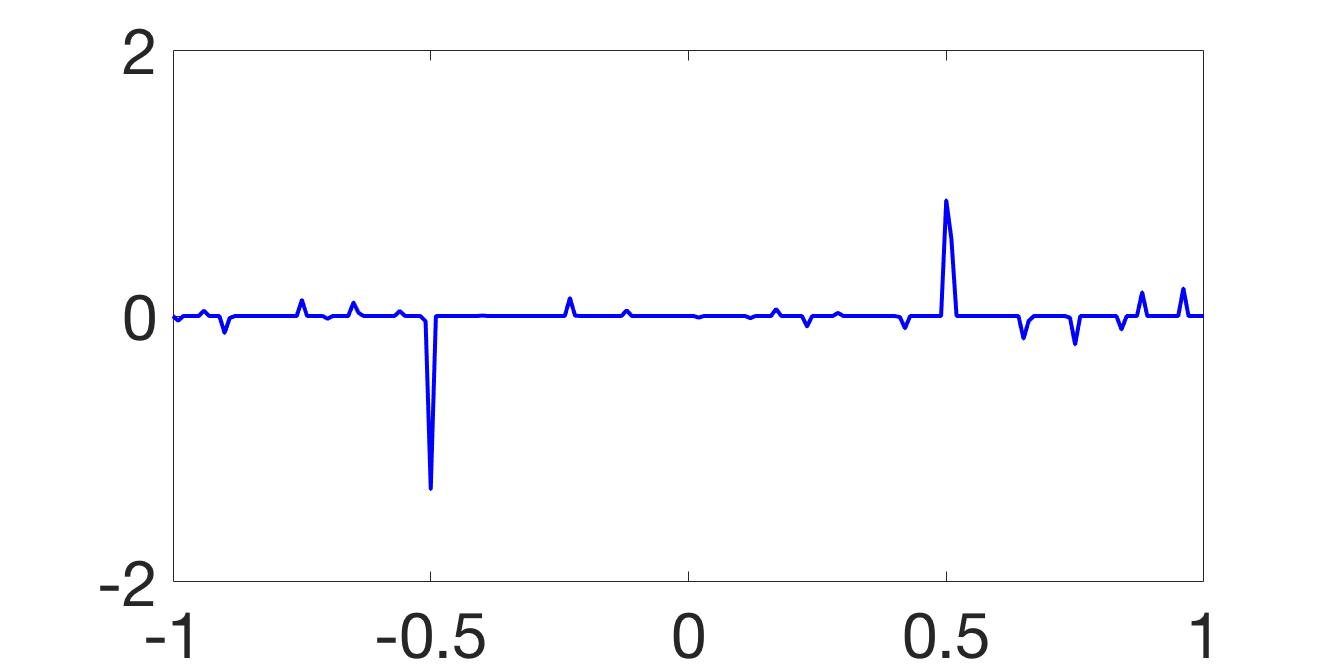}
\includegraphics[width=.32\textwidth]{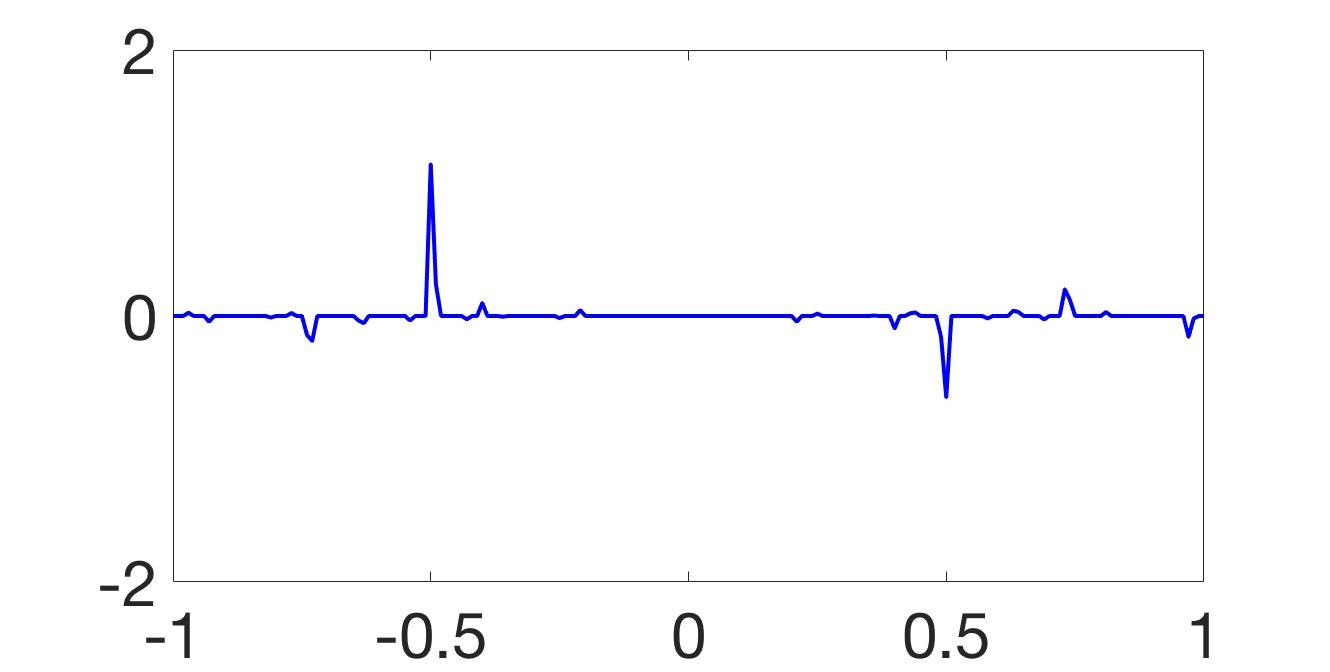}
\includegraphics[width=.32\textwidth]{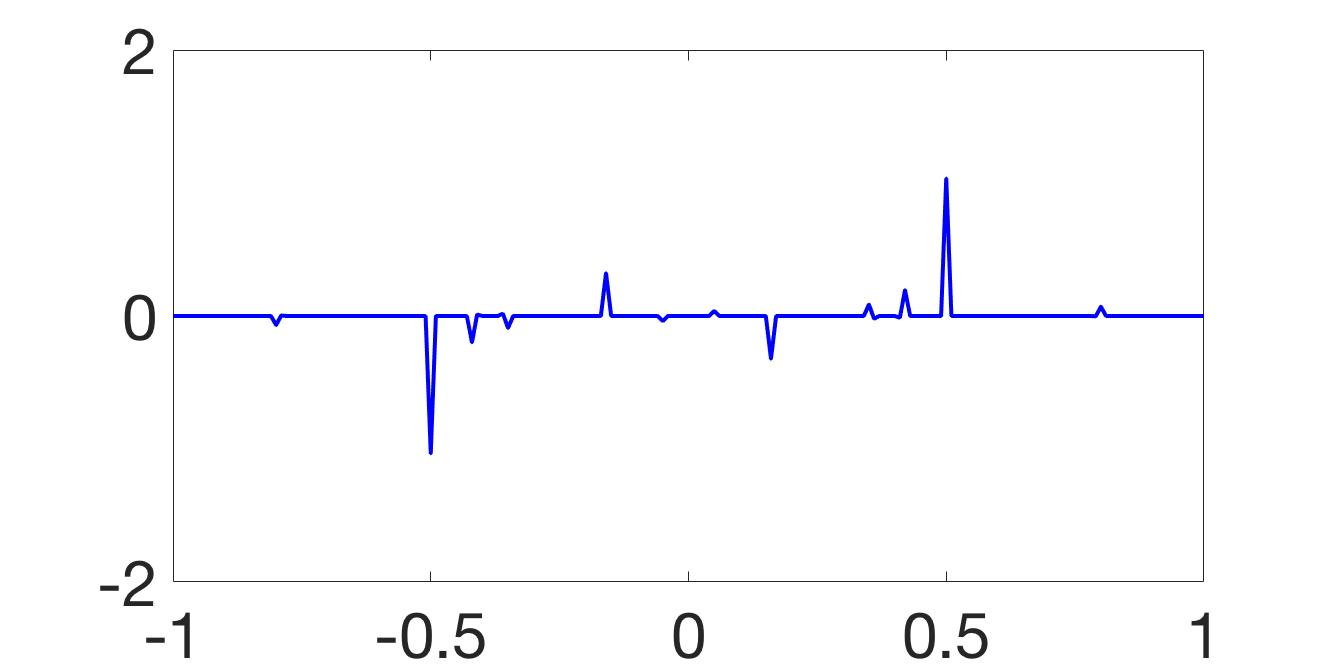}
\caption{Reconstructions with Algorithm \ref{alg:edgel1} from non-uniform Fourier data with zero-mean Gaussian noise with $.02$ standard deviation. (left)$[f_1]$ from logarithmic sampling; (center) $[f_2]$ from quadratic sampling; and (right) $[f_3]$ from jittered sampling. The relative errors left to right are $0.295$, $0.364$, and $0.472$.}
\label{fig:l1_noise}
\end{figure}

Algorithm \ref{alg:edgel1} combines (\ref{eq:model}) with $\ell_1$ regularization, used because of our assumption that the edges are sparse, to construct an approximation to $[f](x)$ at a set of grid points $x_j, j = -J,\ldots,J$.  As in \cite{gelb2017detecting}, we choose $\lambda=.01$ for Algorithm \ref{alg:edgel1}. We also consider the same three examples given there to compare our methods, given in Examples \ref{ex:ex1}, \ref{ex:ex2} and \ref{ex:ex3} below.  The results displayed in Figures \ref{fig:f1_l1}, \ref{fig:f2_l1}, and \ref{fig:f3_l1}, show generally good results for the jittered sampling case, but spurious oscillations appear for both quadratic and logarithmic sampling.  Thus we see that the solutions are not sufficiently sparse, which may have undesirable consequences in downstream processing such as reconstruction. Furthermore, as displayed in Figure \ref{fig:l1_noise}, the results clearly deteriorate when Gaussian noise with mean $0$ and standard deviation $0.02$ is added to the given Fourier data.

\begin{algorithm}[htbp!]
\caption{Reconstruction of $[f](x)$ from non-uniform Fourier data via $\ell_1$ regularization}
\label{alg:edgel1}
\begin{algorithmic}[1]
\STATE Given $2M+1$ Fourier coefficients, $\hat{f}(\lambda_k)$, of a piecewise smooth function, $f$, as in (\ref{eq:fourierdata}).

\STATE Determine $\mathbf{\sigma}$ from Algorithm \ref{alg:cfdesign}.

\STATE Reconstruct the jump function $[f]$ on gridpoints as
\begin{align}\label{eq:l1}
\mathbf{g}^* = \arg\min_{\mathbf{g}} ||\Theta\mathbf{g}-\mathbf{y}||_2+\lambda||\mathbf{g}||_1.
\end{align} 
\end{algorithmic}
\end{algorithm}

\begin{example}
\label{ex:ex1}
\begin{align}
f_1(x) &= \left\{\begin{array}{cc}
\cos(\pi x-\frac{\pi x}{2}\text{ sign}(-x-1/2)) & -1\le x\le0\\
\cos(\frac{5\pi x}{2}+\pi x\text{ sign}(x-1/2)) & 0<x\le1
\end{array}\right.
\end{align}
with
\begin{align}
[f_1](x) &= \left\{\begin{array}{cc}
-\sqrt{2} & x=-1/2\\
\sqrt{2} & x=1/2\\
0 & |x|\neq1/2
\end{array}\right..
\end{align}
\end{example}

\begin{example}
\label{ex:ex2}
\begin{align}
f_2(x) &= \left\{\begin{array}{cc}
-\frac12(1-x^2)^2 & -1\le x\le-1/2\\
\cos(4\pi x) & |x|<1/2\\
(1-x^2)^4 & 1/2\le x\le1
\end{array}\right.
\end{align}
with
\begin{align}
[f_2](x) &= \left\{\begin{array}{cc}
\frac{42}{31} & x=-1/2\\
-\frac{175}{256} & x=1/2\\
0 & |x|\neq1/2
\end{array}\right..
\end{align}
\end{example}

\begin{example}
\label{ex:ex3}
\begin{align}
f_3(x) &= \left\{\begin{array}{cc}
\pi(1-x^2)^2 & 1/2\le |x|\le1\\
-\frac16\sin(6\pi x) & |x|<1/2\\
\end{array}\right.
\end{align}
with
\begin{align}
[f_3](x) &= \left\{\begin{array}{cc}
-\frac{9}{16}\pi & x=-1/2\\
\frac{9}{16}\pi & x=1/2\\
0 & |x|\neq1/2
\end{array}\right..
\end{align}
\end{example}

\begin{figure}[h!]
\centering
\includegraphics[width=.32\textwidth]{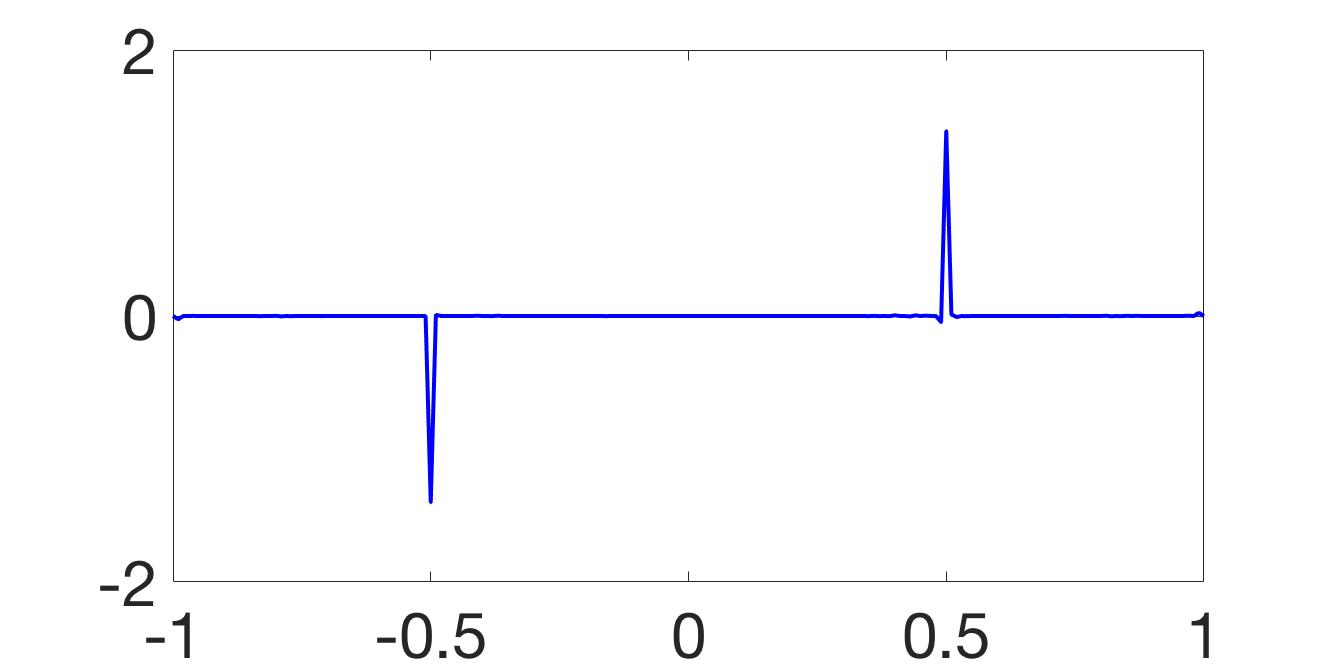}
\includegraphics[width=.32\textwidth]{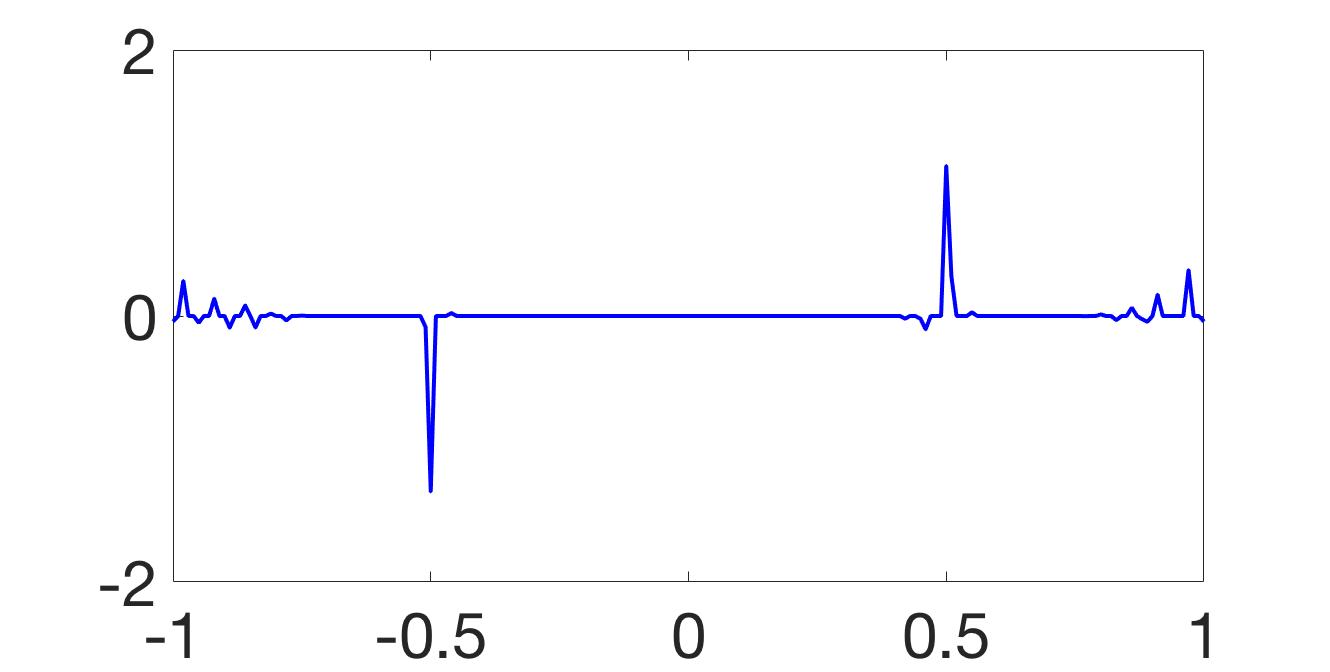}
\includegraphics[width=.32\textwidth]{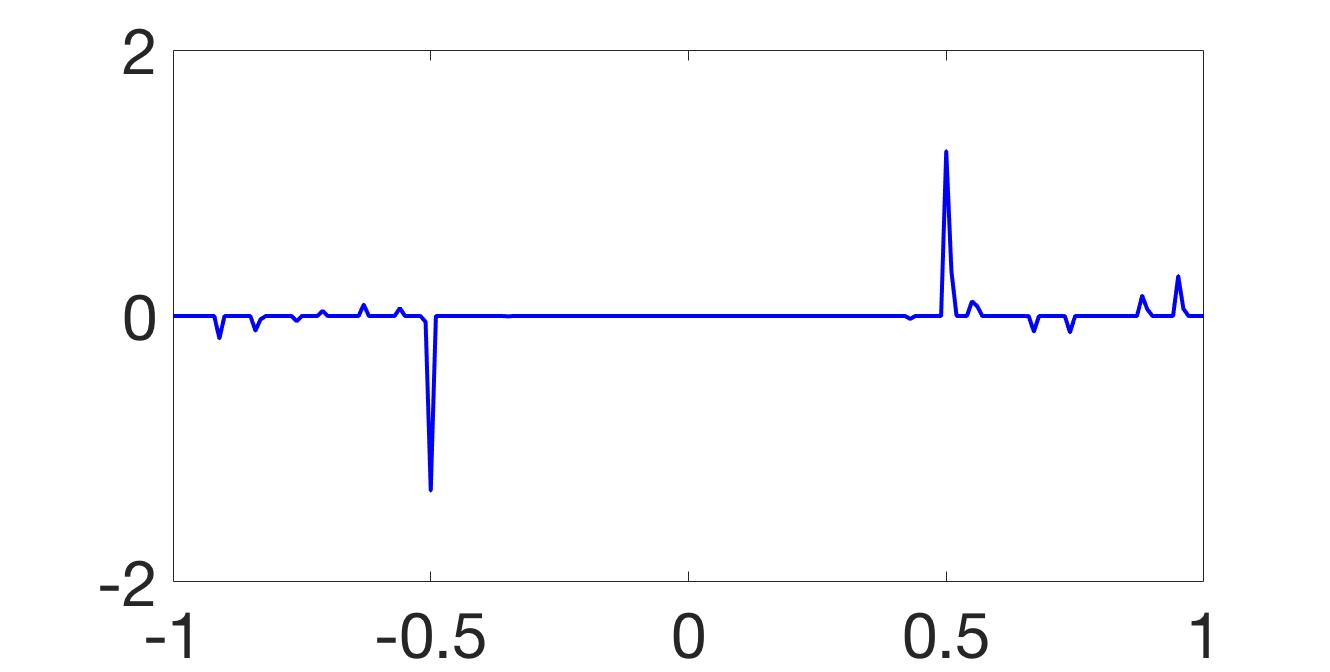}
\caption{$[f_1]$ reconstructed via Algorithm \ref{alg:edgel1} using (left) jittered; (center) quadratic; and (right) logarithmic sampling.}
\label{fig:f1_l1}
\end{figure}

\begin{figure}[h!]
\centering
\includegraphics[width=.32\textwidth]{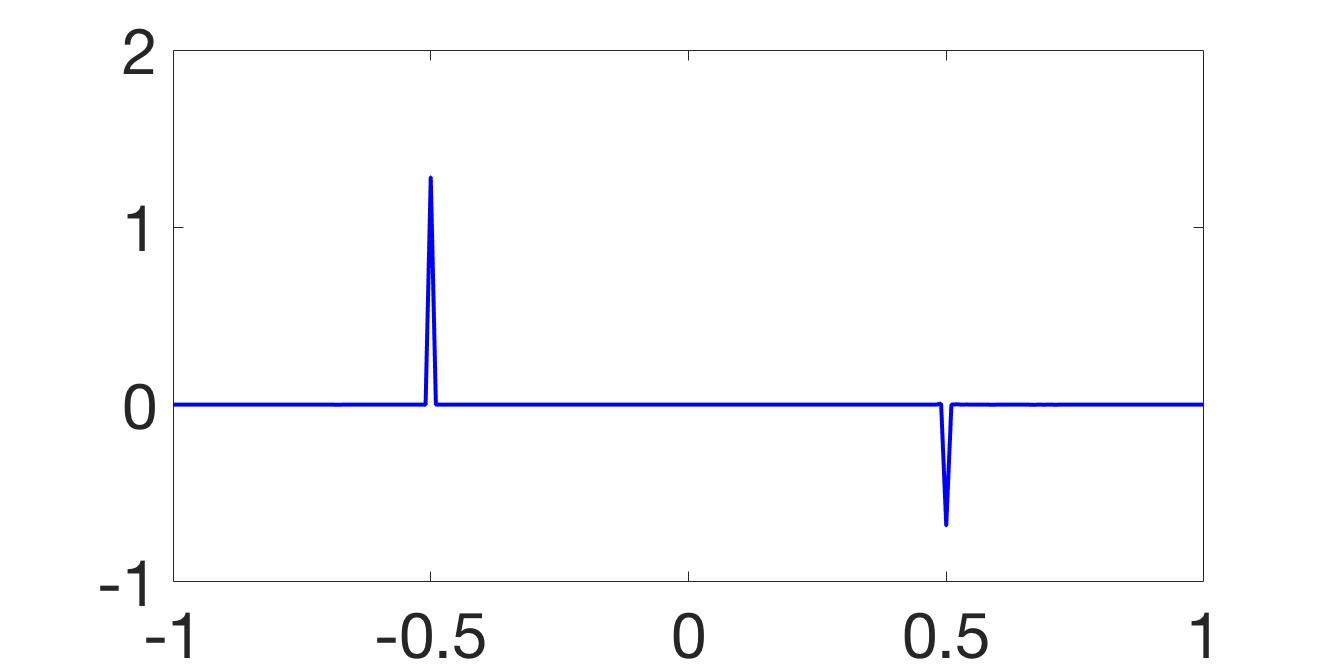}
\includegraphics[width=.32\textwidth]{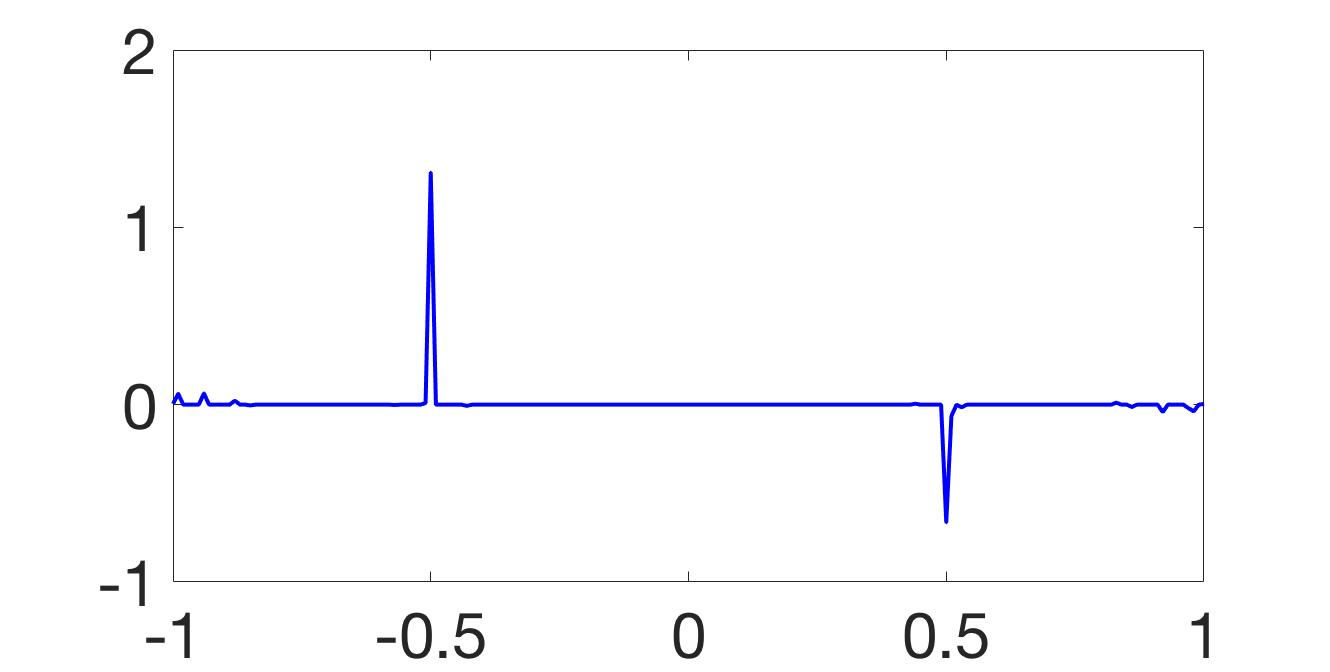}
\includegraphics[width=.32\textwidth]{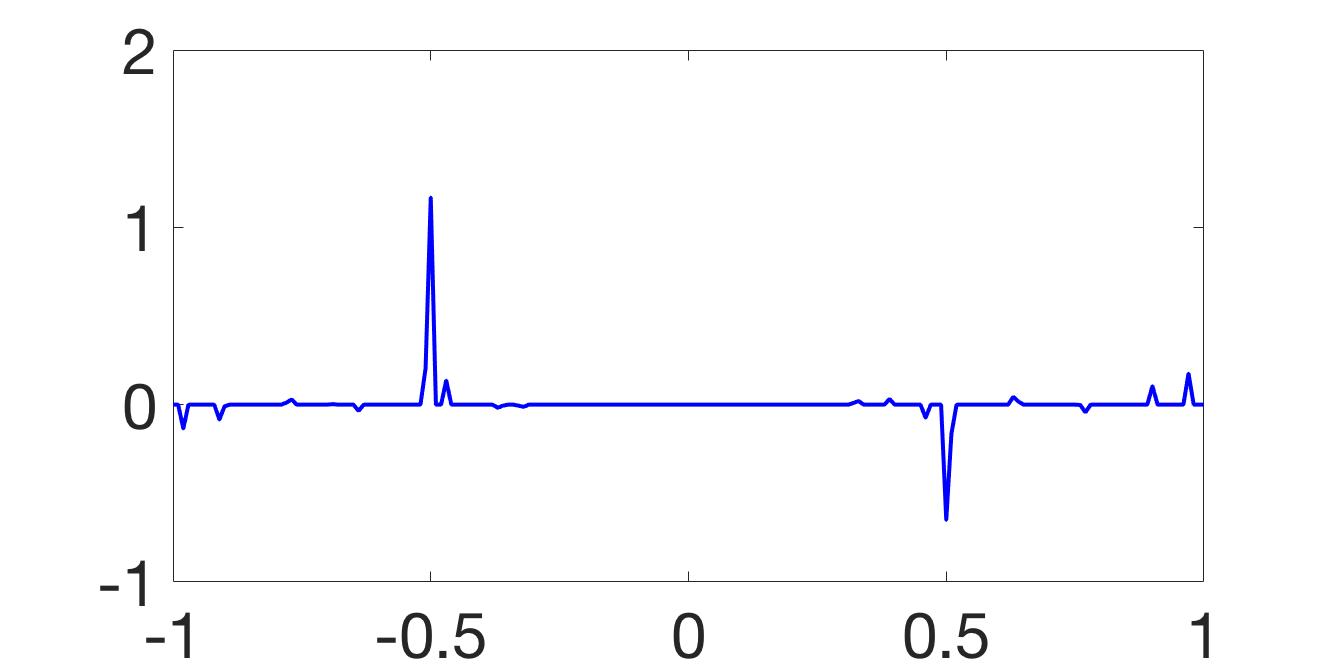}
\caption{$[f_2]$ reconstructed via Algorithm \ref{alg:edgel1} using (left) jittered; (center) quadratic; and (right) logarithmic sampling.}
\label{fig:f2_l1}
\end{figure}

\section{Probabilistic approach}\label{probabilistic}

As Figures \ref{fig:f1_l1}, \ref{fig:f2_l1} and \ref{fig:f3_l1} demonstrate, Algorithm \ref{alg:edgel1} is somewhat effective in recovering edges of piecewise smooth functions from non-uniform Fourier data in cases without noise. However, as Figure \ref{fig:l1_noise} shows, the quality of the edge approximation deteriorates as noise is introduced.  While this suggests that the $\ell_1$ regularization term is no longer mitigating the poor fidelity, it is also apparent that as more noise is introduced, it is more difficult to obtain a truly sparse approximation.  Hence choosing the regularization parameter $\lambda$ a-priori and without oracle knowledge becomes more challenging. As mentioned in the introduction, one way to address this problem is to use the {\em sparse Bayesian learning} (SBL) approach, which we describe below.

Following (\ref{eq:model}), we  now assume that the recovery of the presumably sparse edge map $\mathbf{g}$ can be modeled as 
\begin{align}\label{eq:discretemodel}
\mathbf{y} &= \Theta\mathbf{g}+\mathbf{n}.
\end{align}
Here $\mathbf{y}$ is a transformation of the acquired Fourier data from (\ref{eq:fourierdata}), $\Theta$ is the forward model described in Section \ref{sparsitymodel}, and $\mathbf{n}$ is added noise, assumed zero-mean complex Gaussian with unknown variance $\nu^2$.  Rather than using the strategy in Algorithm \ref{alg:edgel1}, where we find edges by minimizing an explicit $\ell_1$-regularized least squares cost function to enforce sparsity, we now consider the inversion of (\ref{eq:discretemodel}) from a probabilistic perspective. In particular, we use the assumption that the edge function $\mathbf{g}$ is sparse as a prior. There has been increased interest in Bayesian probabilistic approaches to sparse signal recovery (SSR) problems, \cite{babacan2010bayesian,giri2016type,ji2008bayesian,tipping2001sparse,wipf2004sparse}. There are two categories of Bayesian probabilistic methods for SSR that encompass many well-known recovery algorithms in practice, \cite{giri2016type}. The first is type-I, or maximum \textit{a posteriori} (MAP) Bayesian estimation which uses a fixed prior. Type-I includes the most popular methods for SSR from compressed sensing including $\ell_1$ regularization, \cite{chen2001atomic,tibshirani1996regression}, iteratively reweighted $\ell_1$ regularization, \cite{candes2008enhancing}, and iteratively reweighted $\ell_2$ regularization, \cite{chartrand2008iteratively}. On the other hand, Type-II, or evidence maximization Bayesian estimation employs a flexible parametrized prior that is learned from the given data. In addition to the accuracy advantages mentioned in Section \ref{sec:intro}, \cite{giri2016type,ji2008bayesian}, SBL can also provide the added benefit of a full posterior density function, as well as automatic data-driven estimation of hyper-parameters that correspond to the regularization parameter in the MAP method, (\ref{eq:l1}). Below we describe how the SBL approach can be used to improve edge detection given non-uniform Fourier data.

\subsection{$\ell_1$ regularization as a type-I (MAP) estimate}\label{sec:MAP}

We begin by examining $\ell_1$ regularization for SSR from a probabilistic perspective as a type-I (MAP) Bayesian estimate. Because we assume entries of $\mathbf{y}$ are independent, the likelihood of the data given an edge function, $\mathbf{g}$, and added noise variance, $\nu^2$, can be written as the Gaussian likelihood model
\begin{align}\label{eq:glm}
p(\mathbf{y}|\mathbf{g},\nu^2) &= (2\pi\nu^2)^{-(2N+1)/2}\exp\left(-\frac{1}{2\nu^2}||\Theta\mathbf{g}-\mathbf{y}||_2^2\right).
\end{align}
We now formulate the assumption that the edge function $\mathbf{g}$ is sparse by using a sparsity-encouraging prior on $\mathbf{g}$. One widely used sparsity-encouraging prior is the Laplace density function, given by
\begin{align}\label{eq:laplaceprior}
p(\mathbf{g}|\mu) &= \left(\frac{\mu}{2}\right)^{2J+1}\exp\left(-\mu||\mathbf{g}||_1\right).
\end{align}
For given $\mu$, $\nu^2$, using Bayes' theorem yields
\begin{align}\label{eq:MAP}
\mathbf{g}_{MAP} &=\arg\max_\mathbf{g} p(\mathbf{g}|\mathbf{y})= \arg\max_\mathbf{g} p(\mathbf{y}|\mathbf{g},\nu^2)p(\mathbf{g}|\mu) \\&= \arg\min_\mathbf{g} \left\{||\Theta\mathbf{g}-\mathbf{y}||_2^2+2\nu^2\mu||\mathbf{g}||_1\right\},
\end{align}
which is equivalent to (\ref{eq:l1}) where the regularization parameter is $\lambda=2\nu^2\mu$. Thus we see the connection between the typical inversion process via $\ell_1$ regularization as in \cite{candes2006robust,donoho2006compressed} and a MAP approximation to a Bayesian linear regression analysis with a sparsity-encouraging Laplace prior on $\mathbf{g}$, (\ref{eq:MAP}). As it is equivalent to the solution proposed in \cite{gelb2017detecting}, this particular MAP estimate falls victim to the same problems as Algorithm \ref{alg:edgel1}.

In general, there are many sparsity-encouraging priors that can be used to regularize in this context. Functions of this type are sometimes referred to as super-Gaussians as they are characterized by fat tails and a sharp peak at zero. However, as will be seen in what follows, using empirical priors characterized by flexible parameters that are estimated explicitly from the data are more often accurate than type-I estimates.

\subsection{Sparse Bayesian learning (SBL)}\label{sec:SBL}
In the previous section we demonstrated that the $\ell_1$-regularized least squares method of \cite{gelb2017detecting} was equivalent to that of a MAP estimate using a fixed sparsity-encouraging Laplace prior, which may not yield satisfactory results, especially when noise is present as is evident in Figure \ref{fig:l1_noise}.  Further, given the probabilistic formulation in Section \ref{sec:MAP}, we hope to carry Bayesian analysis further and obtain a full posterior distribution on $\mathbf{g}$. This cannot be done using the Laplace prior because it is not conjugate to the Gaussian likelihood model. Conjugate priors, \cite{berger2013statistical}, would allow us to maintain the same functional form for the prior and posterior while only updating the parameters, \cite{babacan2008parameter}. Without conjugacy, the associated Bayesian inference cannot be done in closed form, \cite{bernardo2001bayesian,gelman2014bayesian}. Sparse Bayesian learning (SBL), also referred to as the relevance vector machine (RVM), \cite{tipping2001sparse}, solves this problem by using a hierarchical parametrized prior with similar properties to the Laplace prior. The SBL method derived below closely follows \cite{ji2008bayesian,tipping2001sparse}.

First, define a zero-mean Gaussian prior on each element of $\mathbf{g}$ as
\begin{align}
p(\mathbf{g}|\mathbf{a}) = \prod_{i=-J}^J\mathcal{N}(\mathbf{g}_i|0,\mathbf{a}_i^{-1})
\end{align}
where the hyper-parameter $\mathbf{a}_i$ is the inverse variance of a zero-mean Gaussian density function for each $i=-J,\ldots,J$. This hyper-parameter will be estimated from the data to determine to the spread of this Gaussian and hence the sparsity of $\mathbf{g}$. We consider a non-informative Gamma prior over each element of $\mathbf{a}$, given by
\begin{align*}
p(\mathbf{a}|a,b)=\prod_{i=-J}^J\Gamma(\mathbf{a}_i|a,b).
\end{align*}
We then marginalize over the hyper-parameters $\mathbf{a}$ to obtain the overall prior on $\mathbf{g}$ as
\begin{align}\label{eq:overallprior}
p(\mathbf{g}|a,b) = \prod_{i=-J}^J\int_0^\infty\mathcal{N}(\mathbf{g}_i|0,\mathbf{a}_i^{-1})\Gamma(\mathbf{a}_i|a,b)d\mathbf{a}_i.
\end{align}
Each integral being multiplied in (\ref{eq:overallprior}) is distributed via the Student's $t$-distribution, which, for appropriate $a$ and $b$, is strongly peaked at $\mathbf{g}_i=0$. Therefore this prior favors $\mathbf{g}_i$ being zero, and hence is sparsity-encouraging, similar to the Laplace prior.  A Gamma prior $\Gamma(\beta|c,d)$ is also introduced on $\beta=\frac{1}{\nu^2}$.

The posterior for the edge function $\mathbf{g}$ can be solved for analytically as a multivariate Gaussian distribution
\begin{align}\label{eq:posterior}
p(\mathbf{g}|\mathbf{y},\mathbf{a},\beta) &=\mathcal{N}(\mathbf{g}|\mathbf{m},\Sigma).
\end{align}
with mean and covariance matrix given by
\begin{align}\label{eq:mean}
\mathbf{m} &= \beta\Sigma\Theta^T\mathbf{y},
\end{align}
\begin{align}\label{eq:cov}
\Sigma &= (\beta\Theta^T\Theta+\mathbf{A})^{-1},
\end{align}
where $\mathbf{A}=\text{diag}(\mathbf{a})$, \cite{bishop2006pattern}. If $\mathbf{n}=0$, i.e. there is no added noise, we want to let $\nu^2\rightarrow0$. In \cite{wipf2004sparse}, the authors derive the following expressions for $\mathbf{m}$ and $\Sigma$ in this case as
\begin{align}\label{eq:meannonoise}
\mathbf{m} &= \mathbf{A}^{-1/2}(\Theta\mathbf{A}^{-1/2})^\dagger \mathbf{y},
\end{align}
\begin{align}\label{eq:covnonoise}
\Sigma &= (\mathbf{I}-\mathbf{A}^{-1/2}(\Theta\mathbf{A}^{-1/2})^\dagger\Theta)\mathbf{A}^{-1},
\end{align}
where $\dagger$ is Moore-Penrose pseudoinverse.
We now have the full posterior, so we just need to learn (estimate) hyper-parameters $\mathbf{a}$ and $\beta$ from the given data. By marginalizing over $\mathbf{g}$, the marginal log-likelihood for $\mathbf{a}$ and $\beta$ is
\begin{eqnarray}
\mathcal{L}(\mathbf{a},\beta) &=&\log p(\mathbf{y}|\mathbf{a},\beta)\nonumber\\ 
&=& \log\int p(\mathbf{y}|\mathbf{g},\beta)p(\mathbf{g}|\mathbf{a})d\mathbf{g}\nonumber\\
&=&-\frac12\left((2N+1)\log 2\pi + \log |\mathbf{C}| +\mathbf{y}^t\mathbf{C}^{-1}\mathbf{y}\right),
\label{eq:L}
\end{eqnarray}
with $\mathbf{C}=\beta^{-1}\mathbf{I}+\Theta \mathbf{A}^{-1}\Theta^T$, \cite{bishop2006pattern}. In \cite{ji2008bayesian,tipping2001sparse}, a type-II maximum likelihood approximation is used which utilizes the point estimates for $\mathbf{a}$ and $\beta$ to maximize $\mathcal{L}$ in (\ref{eq:L}), and can be implemented via the {\em expectation-maximization} (EM) algorithm, \cite{dempster1977maximum}, to obtain the update
\begin{align}\label{eq:a}
\mathbf{a}_i^{\text{(new)}}=\frac{\gamma_i}{\mathbf{m}_i^2}
\end{align}
for $i=-J,\ldots,J$, where $\mathbf{m}_i$ is the $i$th posterior mean weight from (\ref{eq:mean}) and $\gamma_i = 1-\mathbf{a}_i\Sigma_{ii}$ with $\Sigma$ from (\ref{eq:cov}). For $\beta$, we obtain the update
\begin{align}\label{eq:beta}
\beta^{\text{(new)}}=\frac{2M+1-\sum_i\gamma_i}{||\mathbf{y}-\Theta\mathbf{m}||_2^2}.
\end{align}
Derivation details for (\ref{eq:a}) and (\ref{eq:beta}) can be found in Appendix A of \cite{tipping2001sparse}.
We have that $\mathbf{a}^{\text{(new)}}$ and $\beta^{\text{(new)}}$ are functions of $\mathbf{m}$ and $\Sigma$, and vise versa, such that the EM algorithm for recovering the posterior for ${\mathbf g}$ iterates between (\ref{eq:mean}) and (\ref{eq:cov}), and (\ref{eq:a}) and (\ref{eq:beta}) until a convergence criterion is satisfied. Due to the properties of the EM algorithm, SBL is globally convergent, meaning each iteration is guaranteed to reduce the cost function until a fixed point is achieved, \cite{wipf2004sparse}. In particular, it has been observed that $\mathbf{a}_i\rightarrow\infty$ for $\mathbf{g}_i\approx0$. Also note that since we seek the point estimates of $\mathbf{a}$ and $\beta$, and not their posterior densities, we do not need to rigorously choose parameters $a$, $b$, $c$, $d$ on the Gamma hyper-priors. Therefore we simply set all of them to $0$, implying uniform hyper-priors (over a logarithmic scale) on $\mathbf{a}$ and $\beta$, \cite{tipping2001sparse}. The resulting algorithm is provided in Algorithm \ref{alg:sbl}.

\begin{algorithm}[htbp!]
\caption{Reconstruction of $[f](x)$ from non-uniform Fourier data via sparse Bayesian learning}
\label{alg:sbl}
\begin{algorithmic}[1]
\STATE Given $2M+1$ Fourier coefficients, $\hat{f}(\lambda_k)$, of a piecewise smooth function, $f$, as in (\ref{eq:fourierdata}).

\STATE Determine $\mathbf{\sigma}$ from Algorithm \ref{alg:cfdesign}.

\STATE Construct $\Theta$ and $\mathbf{y}$.

\STATE Initialize hyper-parameters $\mathbf{a}$ and $\beta$, e.g. $\mathbf{a}_i=1$ for all $i$ or non-negative random initialization.

\STATE Compute $\mathbf{m}$ and $\Sigma$ via (\ref{eq:mean}) and (\ref{eq:cov}).

\STATE Update $\mathbf{a}$ and $\beta$ using (\ref{eq:a}) and (\ref{eq:beta}).

\STATE Repeat steps 5 and 6 until convergence to a fixed point $\mathbf{a}^*$.

\STATE The point estimate for $\mathbf{g}$ is the mean computed with $\mathbf{a}^*$, $\mathbf{m}^*=\mathbf{m}(\mathbf{a}^*)$.
\end{algorithmic}
\end{algorithm}

\begin{figure}[h!]
\centering
\includegraphics[width=.32\textwidth]{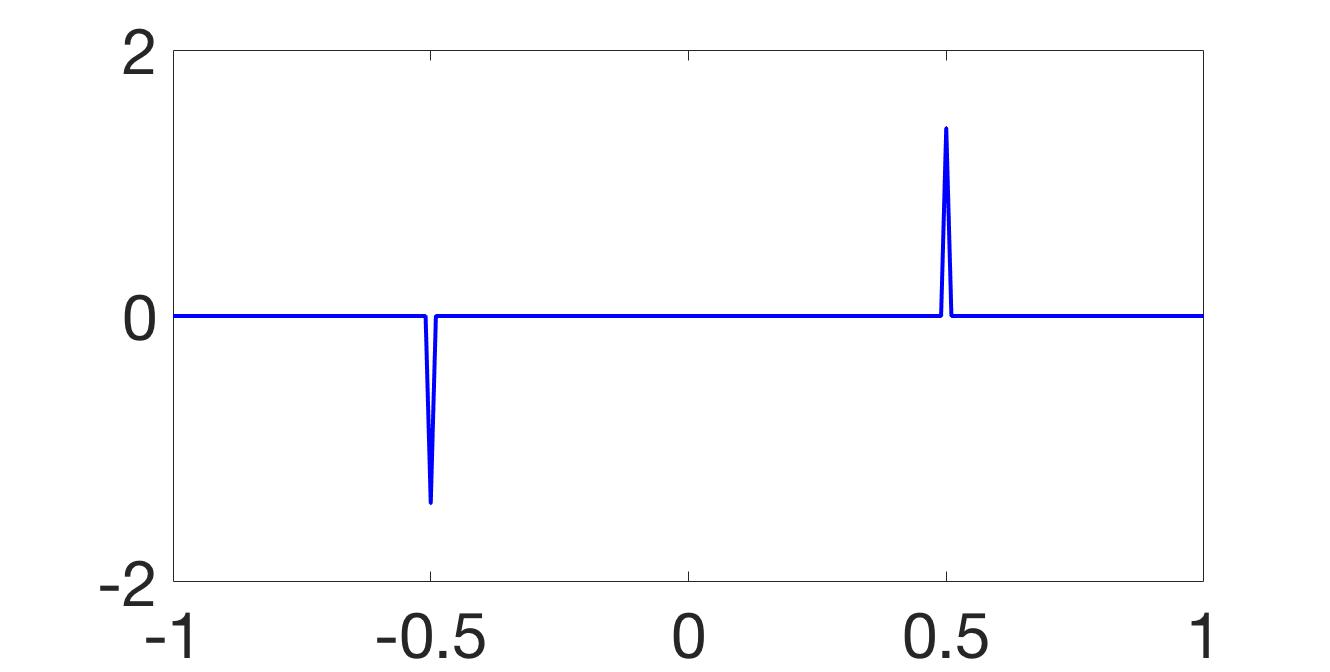}
\includegraphics[width=.32\textwidth]{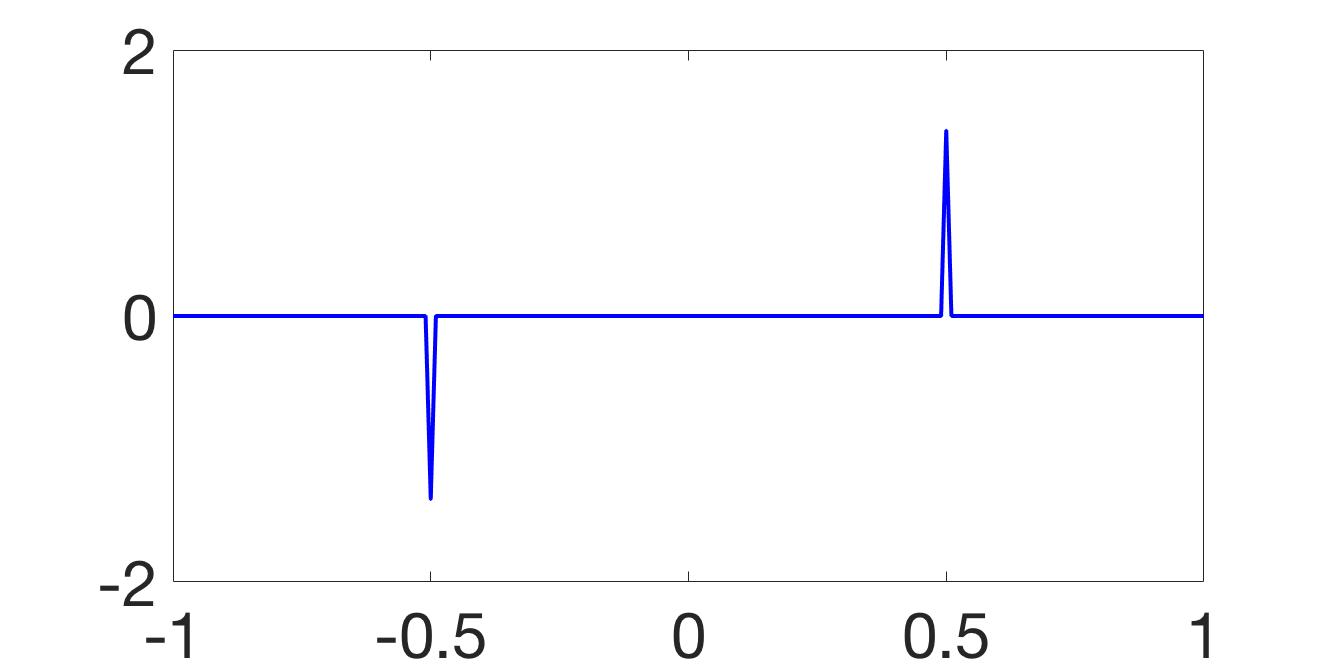}
\includegraphics[width=.32\textwidth]{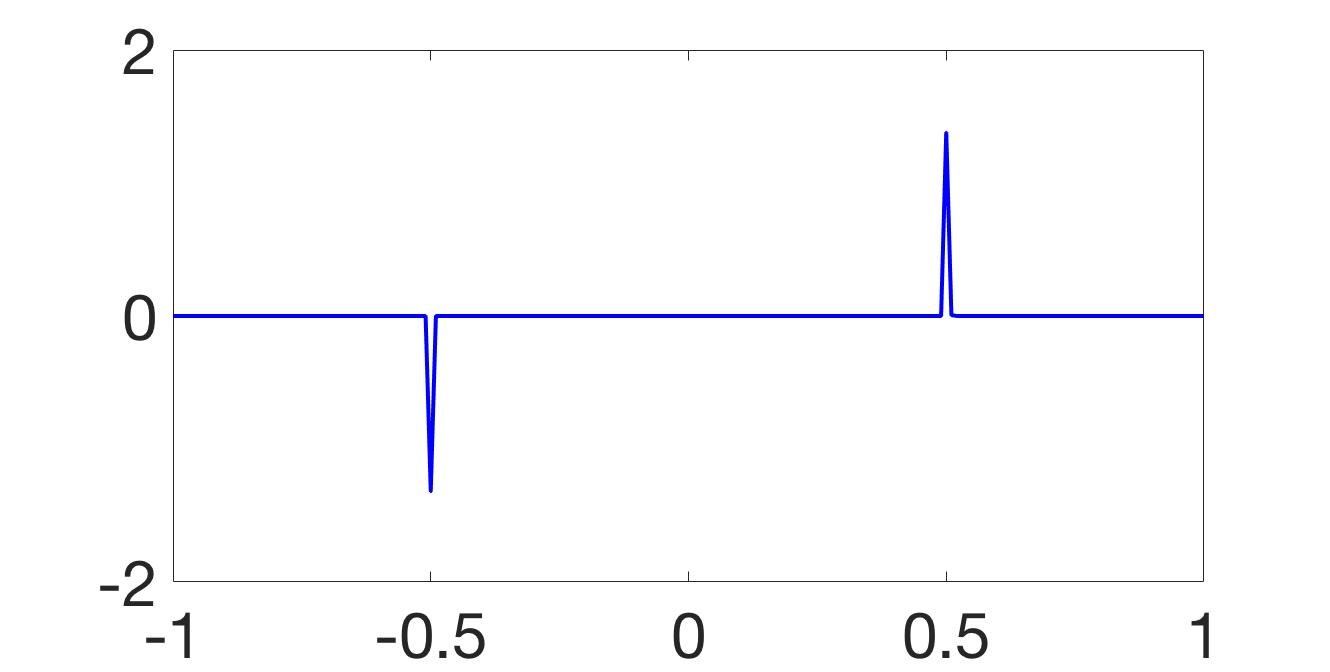}
\caption{$[f_1]$ reconstructed via Algorithm \ref{alg:sbl} using (left) jittered; (center) quadratic; and (right) logarithmic sampling.}
\label{fig:f1_sbl}
\end{figure}

\begin{figure}[h!]
\centering
\includegraphics[width=.32\textwidth]{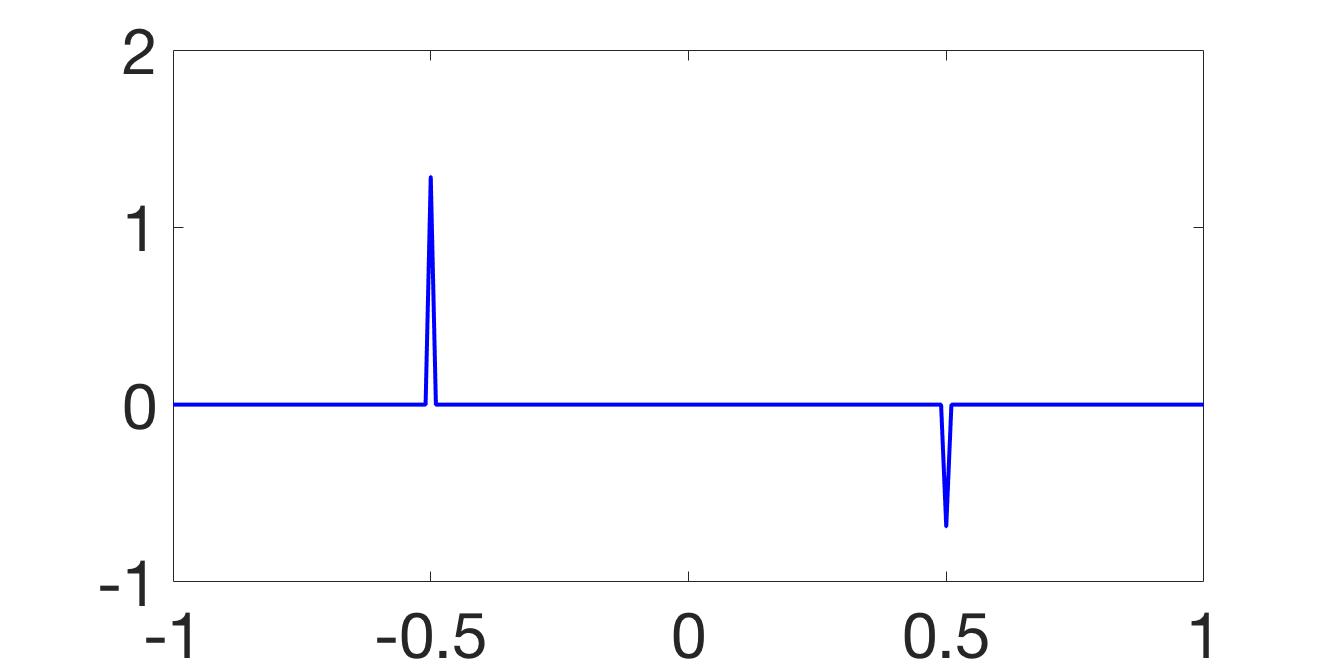}
\includegraphics[width=.32\textwidth]{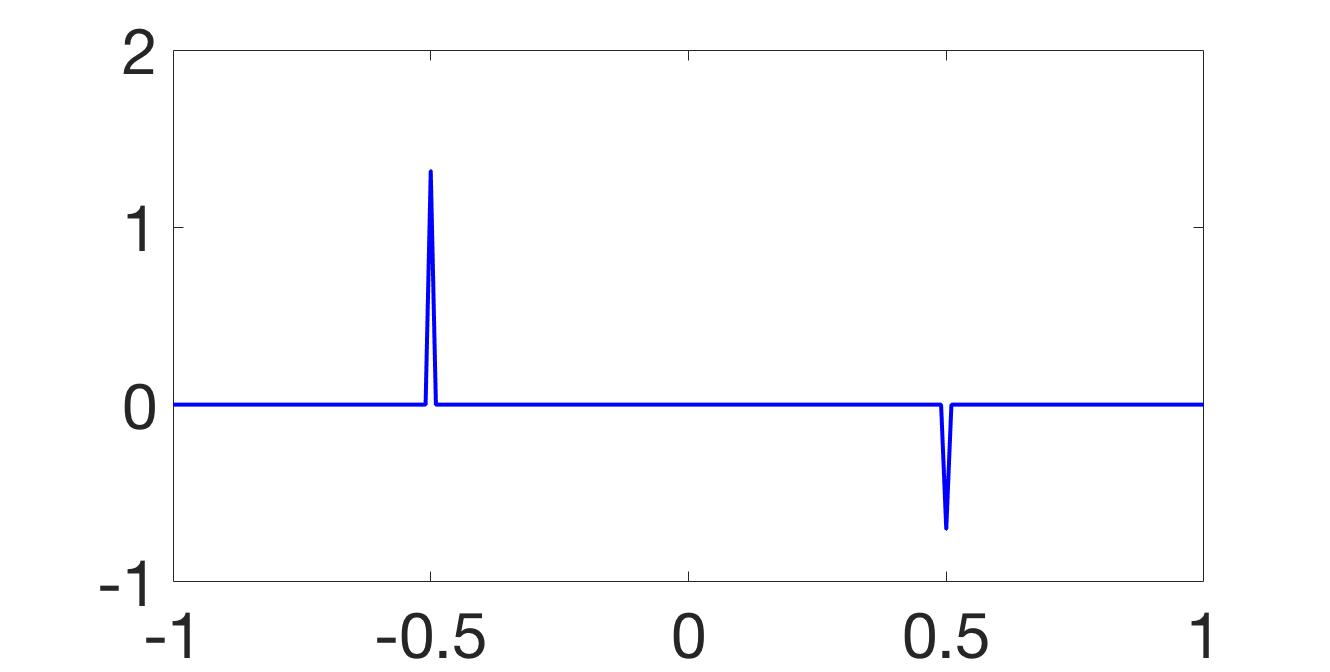}
\includegraphics[width=.32\textwidth]{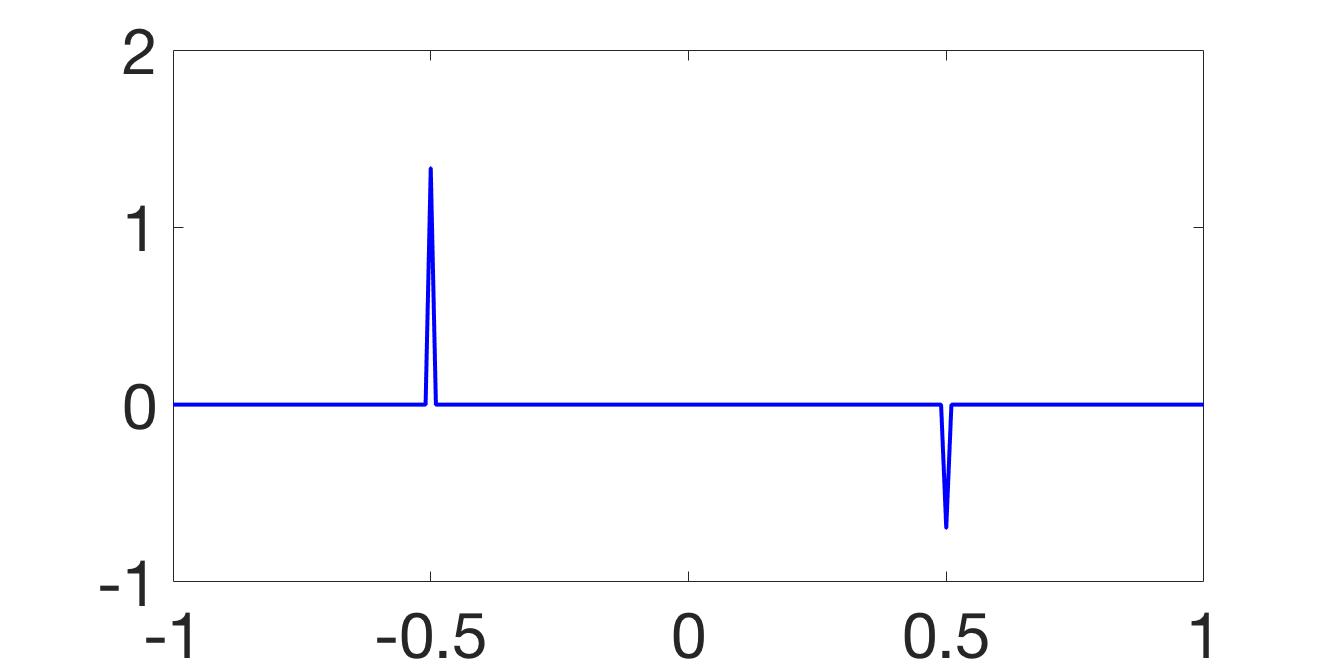}
\caption{$[f_2]$ reconstructed via Algorithm \ref{alg:sbl} using (left) jittered; (center) quadratic; and (right) logarithmic sampling.}
\label{fig:f2_sbl}
\end{figure}

\begin{figure}[h!]
\centering
\includegraphics[width=.32\textwidth]{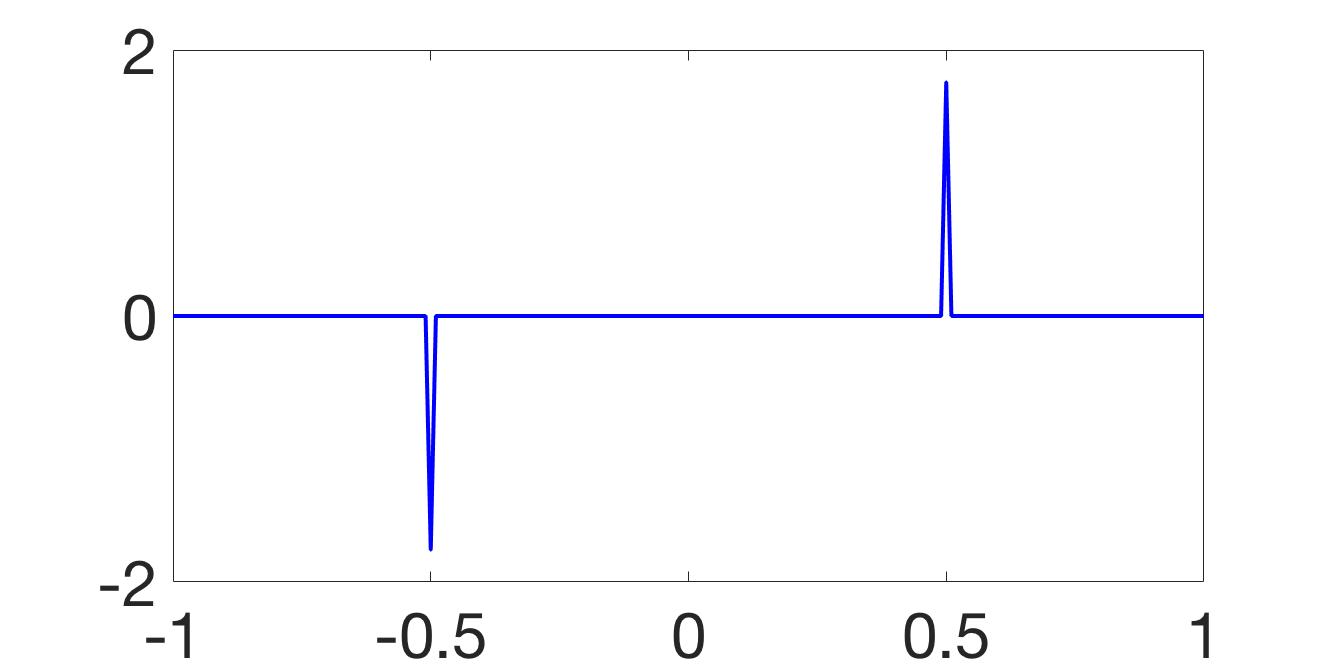}
\includegraphics[width=.32\textwidth]{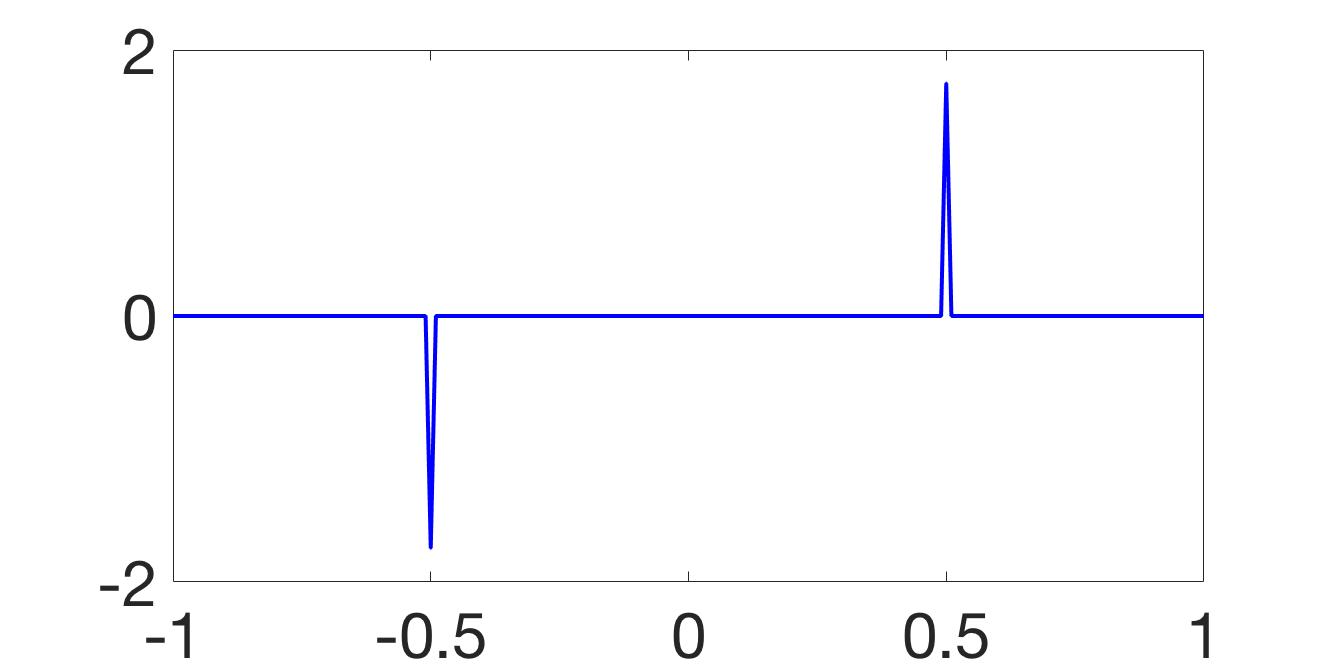}
\includegraphics[width=.32\textwidth]{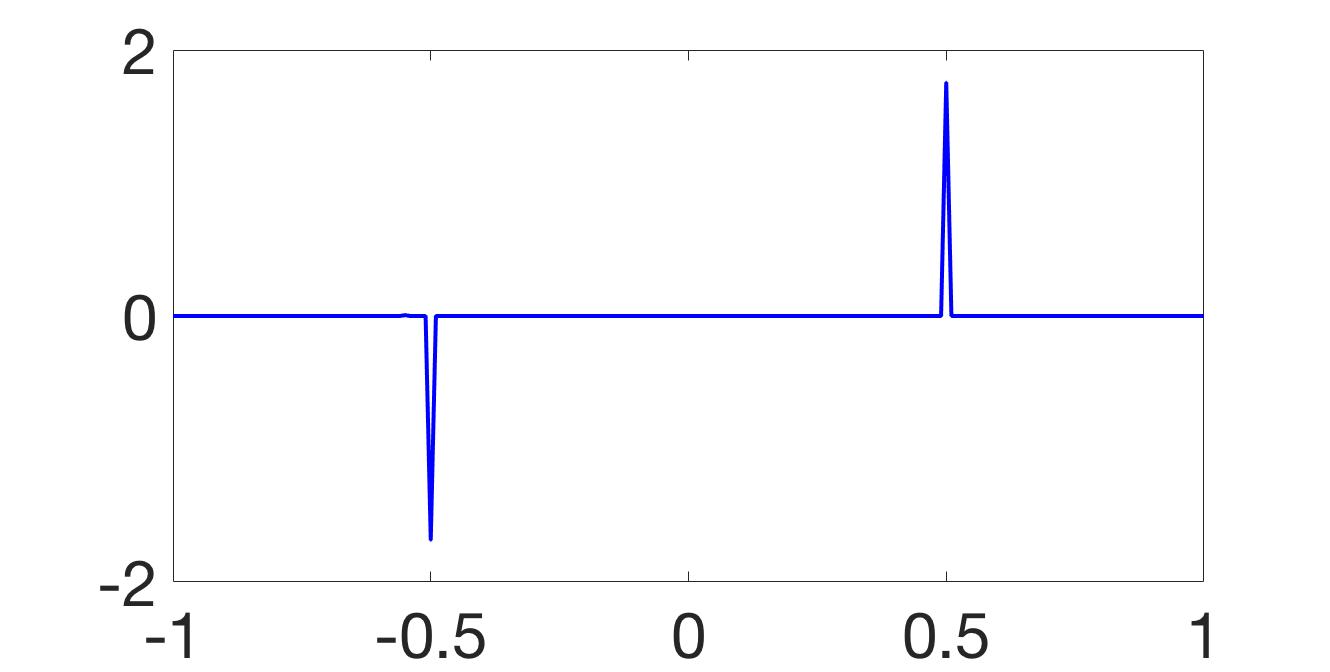}
\caption{$[f_3]$ reconstructed via Algorithm \ref{alg:sbl} using (left) jittered; (center) quadratic; and (right) logarithmic sampling.}
\label{fig:f3_sbl}
\end{figure}

\begin{figure}[h!]
\centering
\includegraphics[width=.32\textwidth]{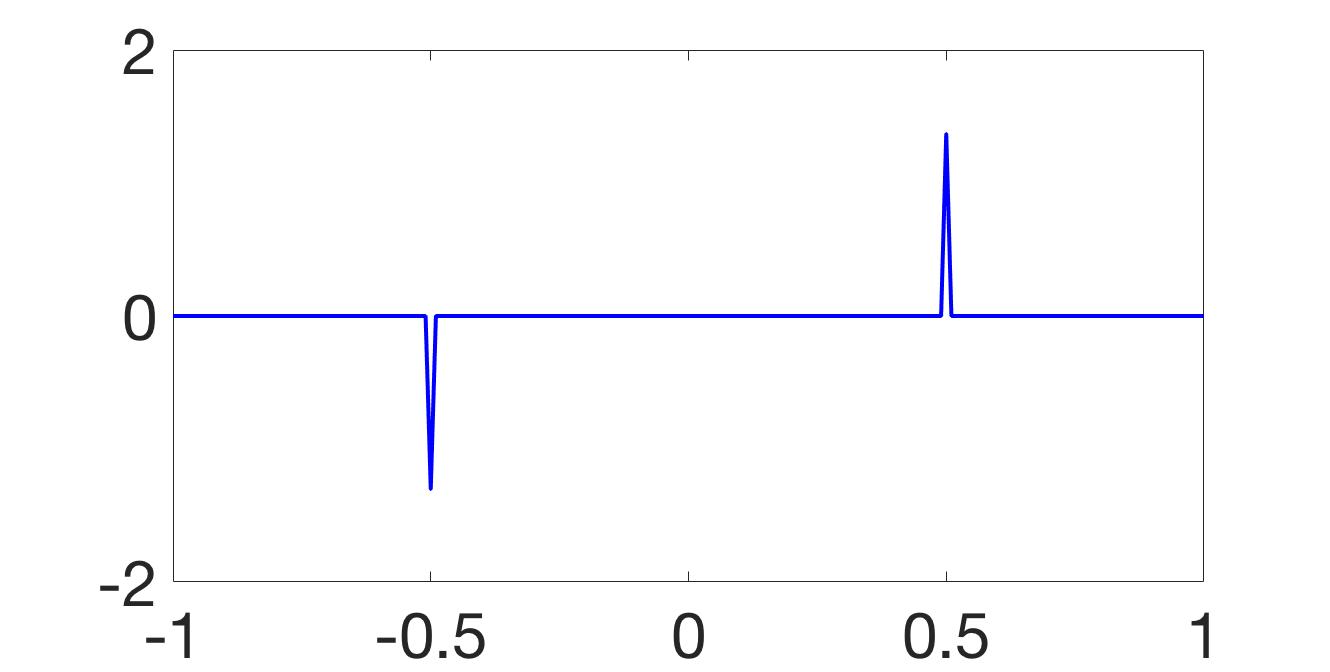}
\includegraphics[width=.32\textwidth]{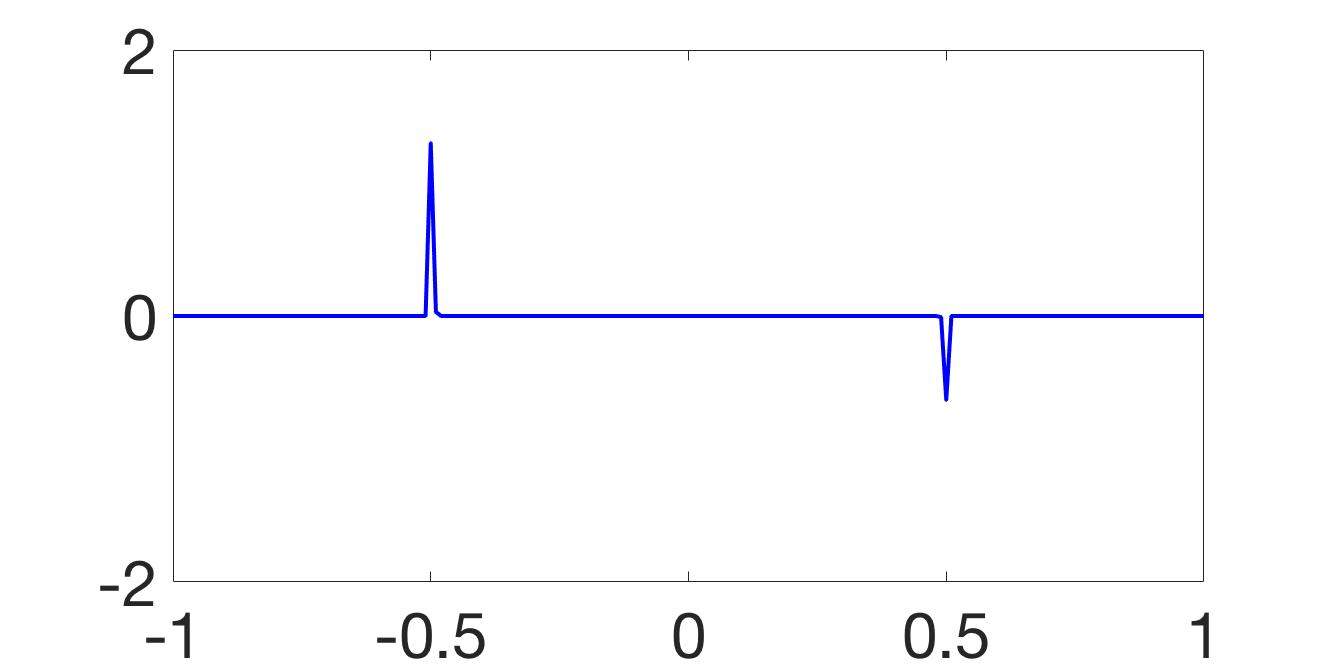}
\includegraphics[width=.32\textwidth]{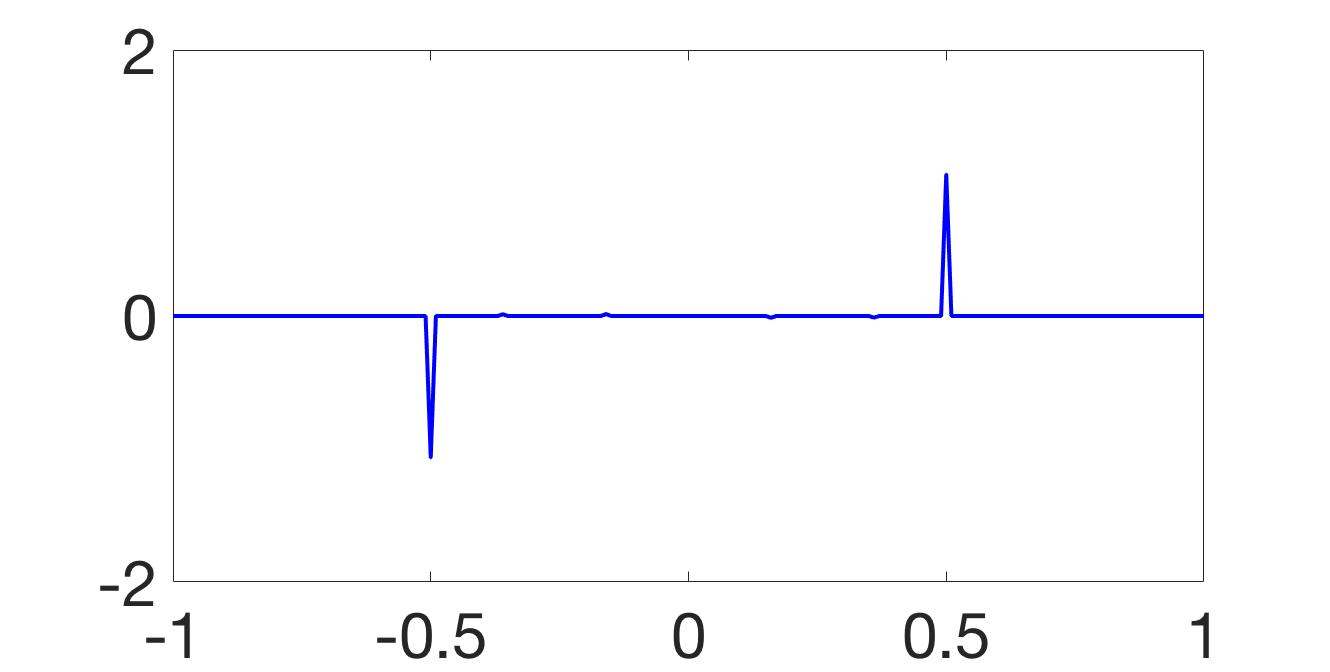}
\caption{Reconstructions with Algorithm \ref{alg:sbl} from non-uniform Fourier data with zero-mean Gaussian noise with $.02$ standard deviation. (left) $[f_1]$ from logarithmic sampling; (center) $[f_2]$ from quadratic sampling; and (right) $[f_3]$ from jittered sampling;. The relative errors left to right are $0.024$, $0.053$, and $0.398$.}
\label{fig:sbl_noise}
\end{figure}

\begin{figure}[h!]
\centering
\includegraphics[width=.32\textwidth]{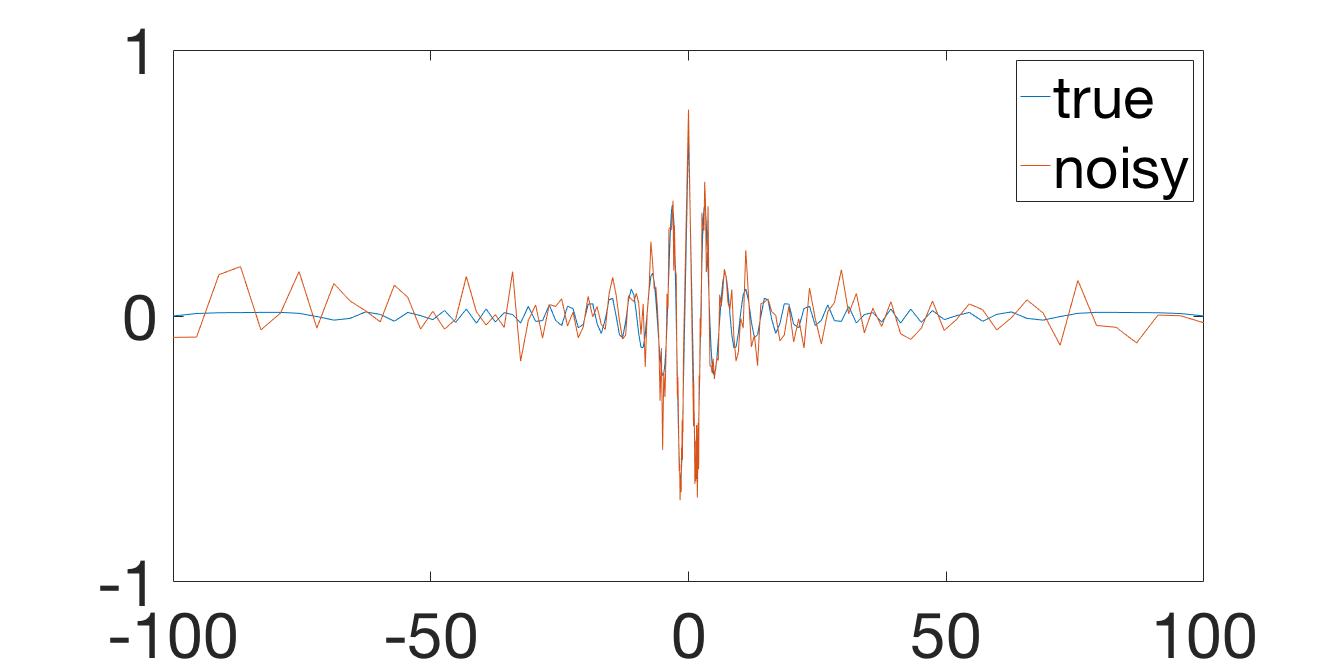}
\includegraphics[width=.32\textwidth]{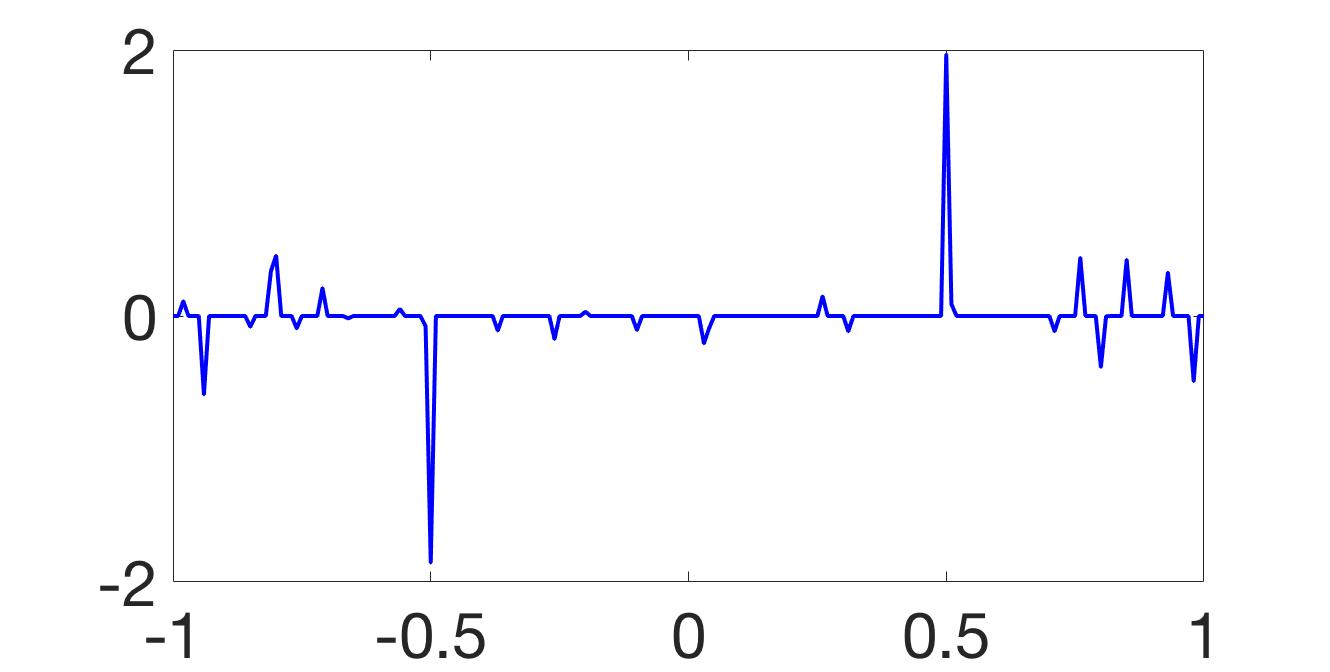}
\includegraphics[width=.32\textwidth]{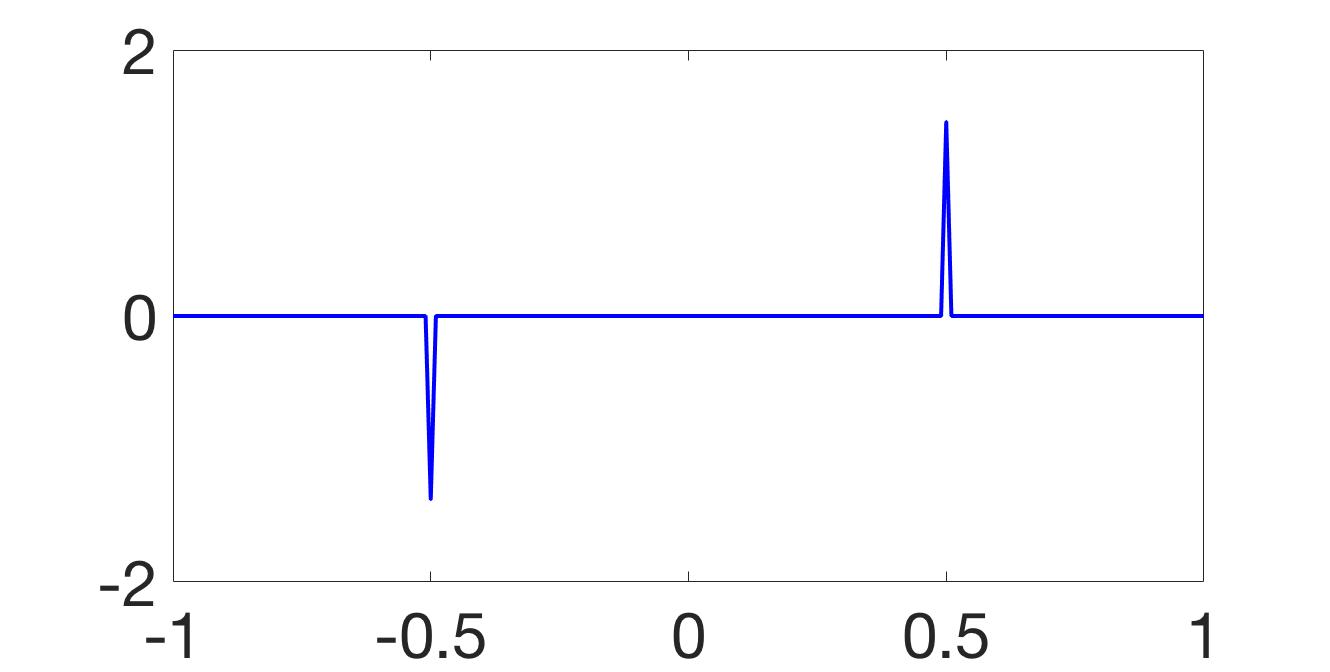}
\caption{Reconstructions of $[f_1]$, given (left) logarithmically sampled non-uniform Fourier data with zero-mean Gaussian noise with $.08$ standard deviation, via (center) Algorithm \ref{alg:edgel1} and (right) Algorithm \ref{alg:sbl}. The relative error in the center is $.753$ and on the right is $.029$.}
\label{fig:sbl_morenoise}
\end{figure}

\begin{figure}[h!]
\centering
\includegraphics[width=.5\textwidth]{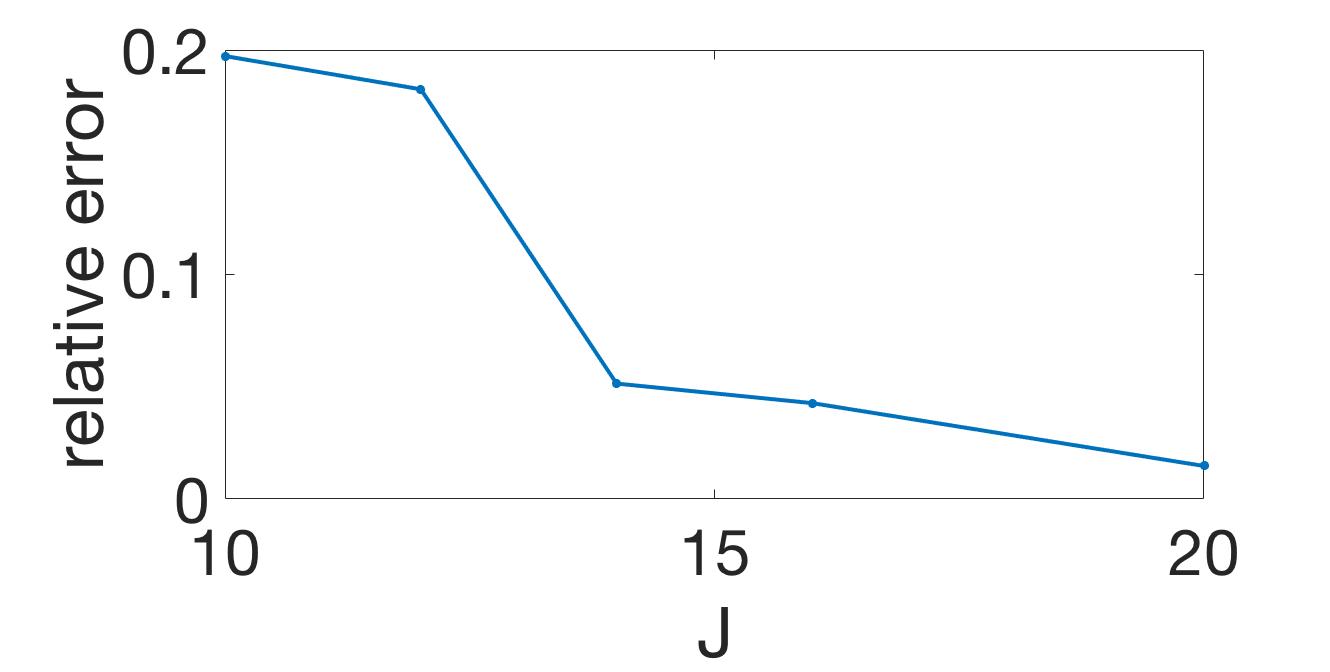}
\caption{Plot of number of grid points $J$ versus relative error. The relative reconstruction error decreases as the number of grid points increases. Each plotted point represents a reconstruction of $[f_3]$ via Algorithm \ref{alg:sbl} from jittered Fourier data with added zero-mean Gaussian noise with $.02$ standard deviation.}
\label{fig:sbl_resolution}
\end{figure}

Figures \ref{fig:f1_sbl}, \ref{fig:f2_sbl}, and \ref{fig:f3_sbl} show the results of Examples \ref{ex:ex1}, \ref{ex:ex2}, and \ref{ex:ex3}, for jittered, quadratic, and logarithmic sampling, respectively. In addition to the visual comparison, Table \ref{table:1} below highlights the dramatic improvements over Algorithm \ref{alg:edgel1} in terms of relative error. When Gaussian noise with mean $0$ and standard deviation $0.02$ is added to the given Fourier data, we retrieve the results shown in Figure \ref{fig:sbl_noise}. Notice the significant improvement over Figure \ref{fig:l1_noise} both visually in the ease of identifying true jump locations and in terms of relative error which focuses more on accurately estimating the height of the jump. Figure \ref{fig:sbl_morenoise} shows another example using even more noise.
Finally, Figure \ref{fig:sbl_resolution} demonstrates that the error depends on the ratio of jumps to grid points in the reconstruction domain. In that example, we see that as the number of grid points increases, the relative reconstruction error decreases as well.

\begin{table}
\caption{Comparison of relative errors from Algorithms \ref{alg:edgel1} and \ref{alg:sbl} over example functions $f_1$, $f_2$, and $f_3$ using jittered, quadratic, and logarithmic sampling in the noise-free case.}
\label{table:1}
\begin{center}
 \begin{tabular}{|c|c|c|c|} 
 \hline
  &  jittered &  quadratic &  logarithmic \\ 
 \hline
 $f_1$ with Alg. \ref{alg:edgel1} & 0.0381 & 0.2475 & 0.3072   \\ 
 \hline
 $f_1$ with Alg. \ref{alg:sbl} & 0.0018 & 0.0229 & 0.0515 \\
 \hline
  & & & \\
  \hline
 $f_2$ with Alg. \ref{alg:edgel1} & 0.0485 & 0.0856 & 0.2978\\
 \hline
 $f_2$ with Alg. \ref{alg:sbl} &  0.0466 & 0.0277 & 0.0158  \\
 \hline
   & & & \\
  \hline
 $f_3$ with Alg. \ref{alg:edgel1} & 0.0156 & 0.1513 & 0.2884\\
 \hline
 $f_3$ with Alg. \ref{alg:sbl} & $5.4364\times10^{-4}$ & 0.011 & 0.034\\
 \hline
\end{tabular}
\end{center}
\end{table}


We note that it is possible to further improve these results by non-linearly post-processing the results of Algorithm \ref{alg:sbl}. For example, the algorithm can run  multiple times with different concentration factors. Points that differ in sign between the two runs indicate an oscillatory response -- not a true edge -- and are set to zero. Assuming an appropriate choice for each of the concentration factors, this will only affect falsely identified jumps, further refining the accuracy result in the noisy case.

\begin{figure}[h!]
\centering
\includegraphics[width=.5\textwidth]{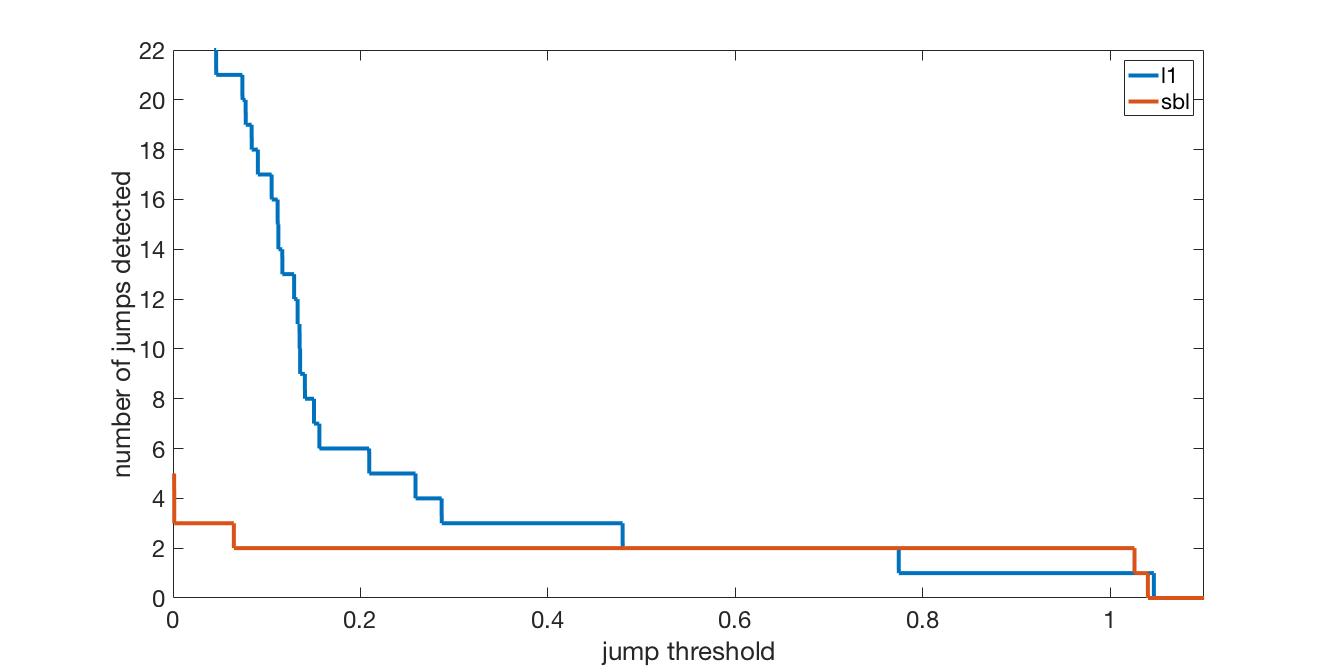}
\caption{Detection comparison of Algorithms \ref{alg:edgel1} and \ref{alg:sbl} based on number of jumps detected for a given threshold. This plot is derived from the reconstruction of $[f_1]$ from logarithmic sampling with zero-mean Gaussian noise with $.02$ standard deviation.}
\label{fig:threshold}
\end{figure}

To summarize the superior performance of Algorithm \ref{alg:sbl} to Algorithm \ref{alg:edgel1}, we consider a detection scenario where we would like to be able to distinguish true jumps from false jumps that are either artifacts of the edge detection process or misclassified due to noise. In this scenario, a grid point will be classified as a jump if the value at that grid point is above a certain threshold. For a particular example, Figure \ref{fig:threshold} shows the number of jumps identified for many thresholds. We see that Algorithm \ref{alg:sbl} correctly identifies the two correct jumps (we verified they are indeed at the correct locations) and only those two for a much wider range of threshold values than the $\ell_1$ edge detection method.

\subsection{Discussion}\label{Discussion}
Our numerical experiments clearly demonstrate that Algorithm \ref{alg:sbl} yields better accuracy than $\ell_1$ regularization (Algorithm \ref{alg:edgel1}) for both the noiseless and noisy cases.  The SBL results are better both in terms of damping down false edges as well as more accurately estimating jump height. The Bayesian formulation also provides advantages over other formulations not readily apparent in the figures above. Namely it allows for automatic estimation of model parameters, \cite{babacan2010bayesian}. Choosing parameters in $\ell_1$-regularized formulations is problem dependent, and often by trial and error, sometimes requiring fine tuning. In SBL, the analog of the regularization parameter $\lambda$ as in (\ref{eq:l1}) are the hyper-parameters $\mathbf{a}$ and $\beta$, \cite{wipf2004sparse}, both of which are estimated directly from the data. This removes user input from the process, which is a common complaint from practitioners regarding $\ell_1$ regularization, since it means that the methods are neither automated nor robust to perturbations in the data. That being said, in very noisy scenarios $\beta$ can still be chosen beforehand instead of estimated, in particular when the point estimate is really the goal. Also not apparent in our numerical experiments is the fact that we have obtained a full density on the edge map $\mathbf{g}$, where only a point estimate is available when using the type-I MAP method. This will be particularly useful in downstream processes where a probability distribution on each gridpoint's value is required. Moreover, because of our accurate estimation of $\Theta$ in \eqref{eq:model} with regard to the non-uniform Fourier data acquisition, we are better able to capture different magnitudes of edges, which can be useful in classifying a variety of targets.  As was pointed out in \cite{martinez2014edge}, determining an edge map directly from the (non-uniform) Fourier data, as opposed to first reconstructing an image (e.g.~via non-uniform FFT) and then using an image based edge detection method (such as Canny edge detection), always yields improved fidelity, especially when the data are highly non-uniform or noise is present.  This is because important edge information is not lost (filtered) in the image reconstruction process.

SBL is not without its own problems, however. While the iterative algorithm described above has been demonstrated to yield highly accurate sparse solutions, \cite{giri2016type}, with theoretical guarantees, \cite{rao2006comparing,wipf2004sparse,wipf2005norm}, Algorithm \ref{alg:sbl} requires the inversion of the $(2J+1)\times (2J+1)$ covariance matrix $\Sigma$ in (\ref{eq:cov}), an $\mathcal{O}((2J+1)^3)$ operation, at each step. This can be slow as the signal size grows. To mitigate this problem, a workaround was suggested in \cite{wipf2004sparse} using the following simplification:
\begin{eqnarray}
\Sigma &=&(\beta\Theta^T\Theta+\mathbf{A})^{-1}\nonumber\\
&=&\mathbf{A}^{-1}-\mathbf{A}^{-1}\Theta^T(\beta^{-1}\mathbf{I}+\Theta\mathbf{A}^{-1}\Theta^T)^{-1}\Theta\mathbf{A}^{-1}\nonumber\\
&=&\mathbf{A}^{-1} - \mathbf{A}^{-1}\Theta^T\mathbf{C}^{-1}\Theta\mathbf{A}^{-1}.
\end{eqnarray}
Now the computation of $\Sigma$ only requires inverting the $(2N+1)\times(2N+1)$ matrix $\mathbf{C}$, reducing the complexity to $\mathcal{O}((2N+1)^3)$, which is much faster for $N< J$. Other fast algorithms to consider for optimal implementation include \cite{faul2002analysis,ji2008bayesian,tipping2003fast}.

We also want to address other type-II methods. Another hierarchical parametrized prior-based estimation technique to consider is the so-called type-II $\ell_1$ method, \cite{babacan2010bayesian,giri2016type}. Effectively, this prior uses the same hierarchical structure but slightly different update rules than those prescribed in (\ref{eq:a}) and (\ref{eq:beta}). In \cite{babacan2010bayesian}, empirical evidence supports the idea that using the same number of measurements, type-II $\ell_1$ consistently produced lower reconstruction error than SBL. However, in \cite{giri2016type} it was empirically shown that using the same number of measurements, for a signal with the same number of nonzero entries, SBL consistently had a higher probability of successfully reconstructing the signal. Our goal in this current investigation is to develop an algorithm for edge detection using SBL and provide some numerical results. The alternative constructions described here may be effective modifications to our basic algorithm, and will be explored in future work.

\section{Applications of edge detection via SBL}\label{sec:applications}
As noted in the introduction, edge detection is not only useful in and of itself, but can be used for downstream processing such as signal recovery and image reconstruction, which can in turn be used for classification, target identification, and change detection. There are two main advantages to employing Algorithm \ref{alg:sbl} over classical forms of $\ell_1$ regularization.  First, due to the construction of $\Theta$ in \cite{gelb2017detecting}, we achieve better accuracy to the {\em magnitude} of the edges in the sparse signal recovery.  Second, while Algorithm \ref{alg:sbl} is designed to recover the mean of the solution, it is straightforward to recover the full posterior distribution described in (\ref{eq:posterior}), (\ref{eq:mean}), and (\ref{eq:cov}).  In what follows we demonstrate how SBL improves image reconstruction.  Other applications using our SBL approach will be the subject of future investigations.

\begin{figure}[h!]
\centering
\includegraphics[width=.32\textwidth]{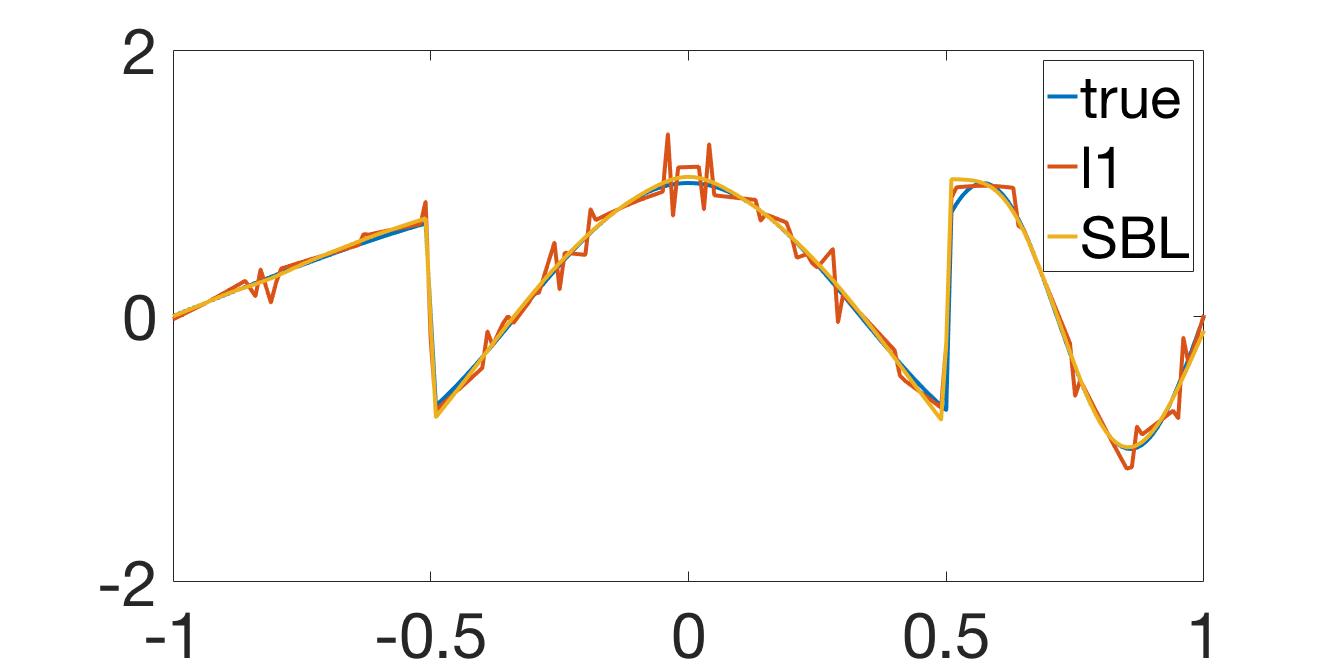}
\includegraphics[width=.32\textwidth]{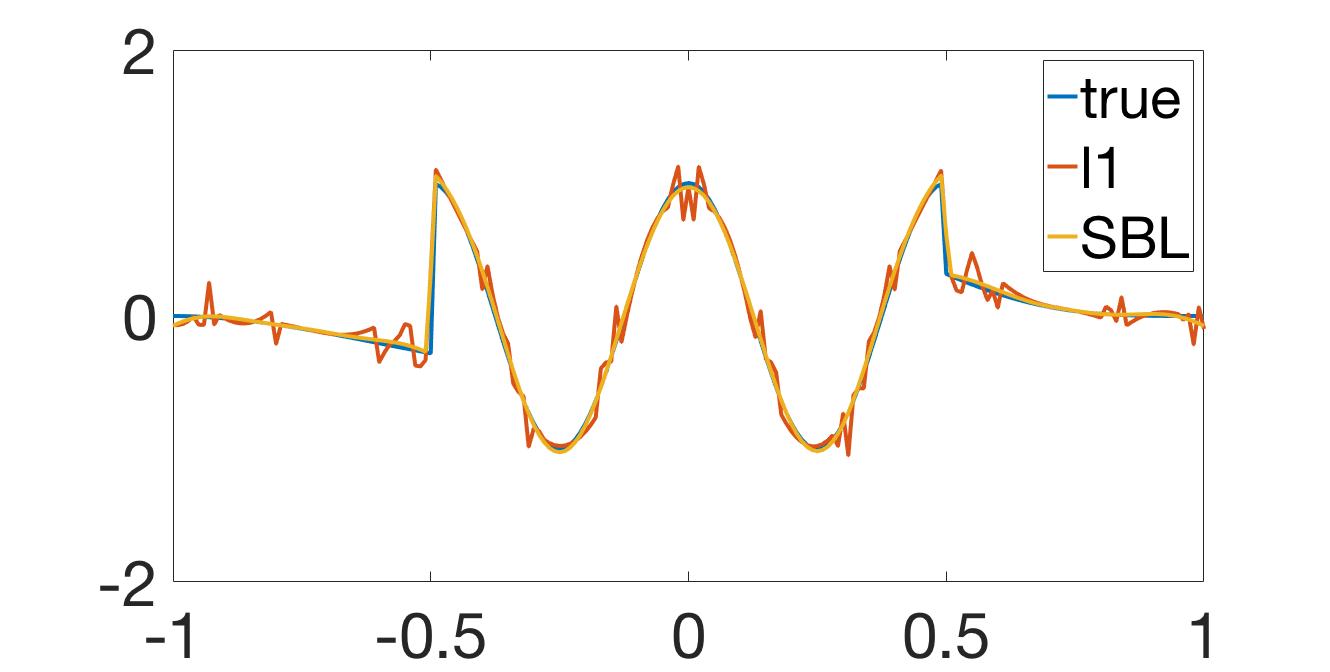}
\includegraphics[width=.32\textwidth]{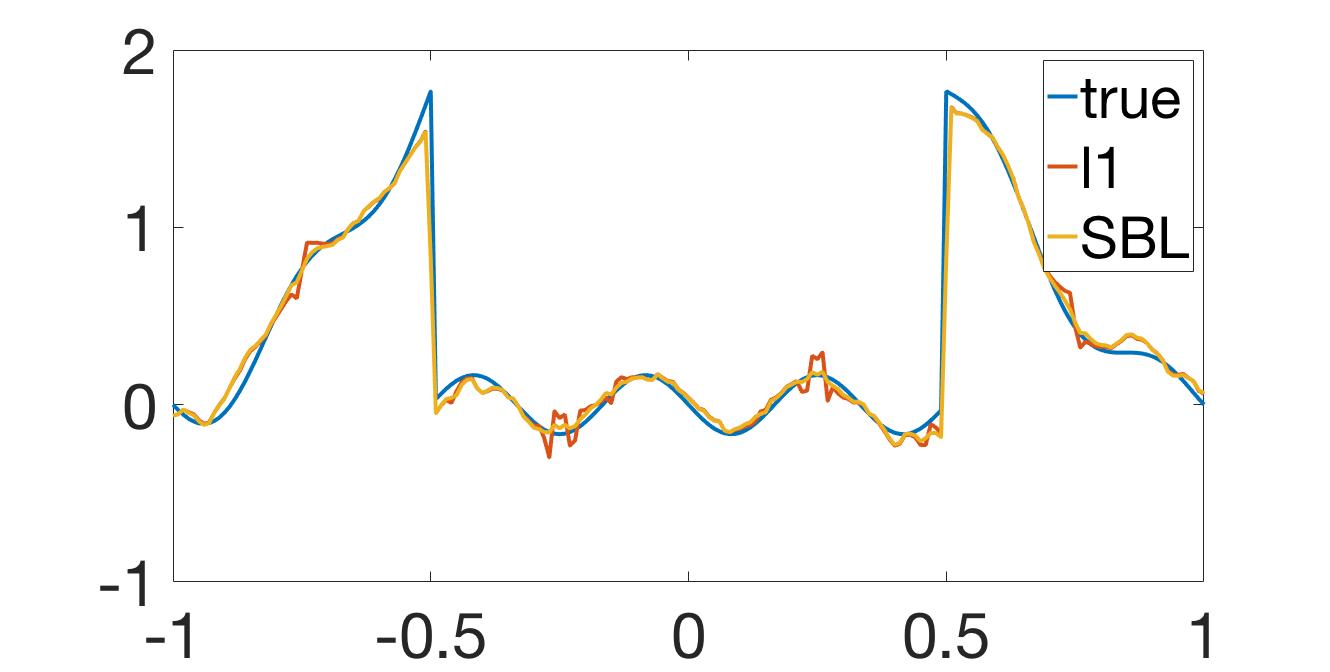}
\includegraphics[width=.32\textwidth]{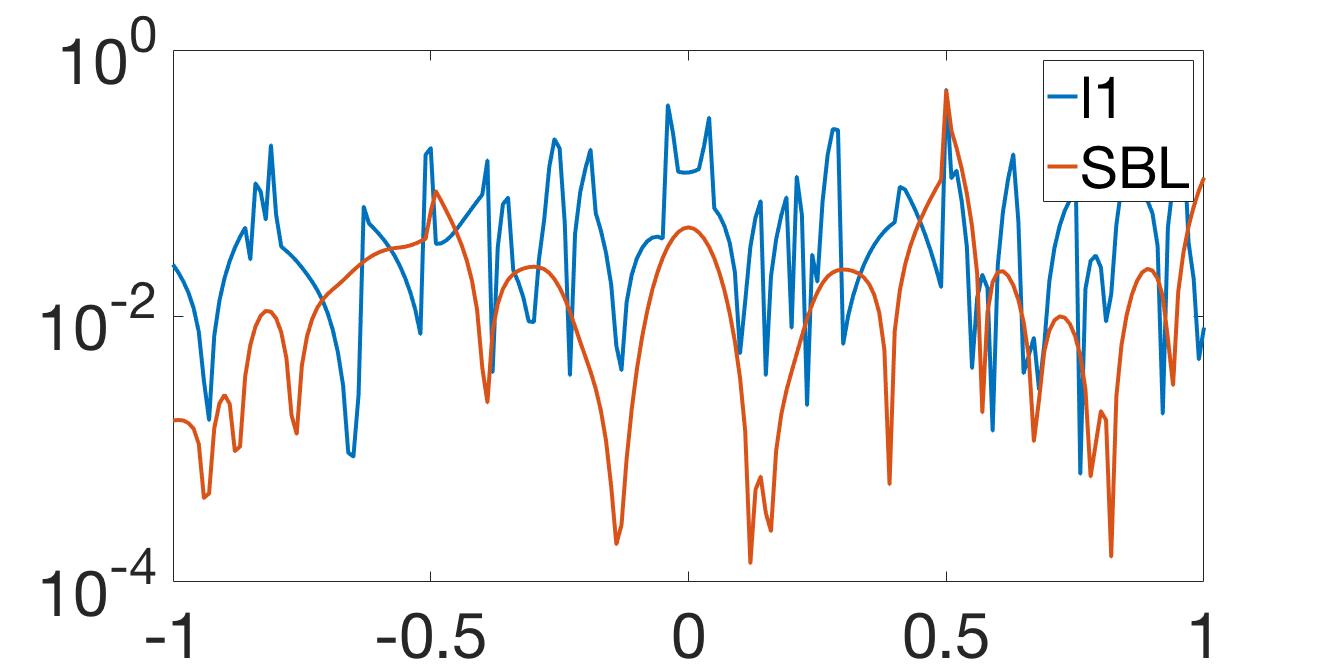}
\includegraphics[width=.32\textwidth]{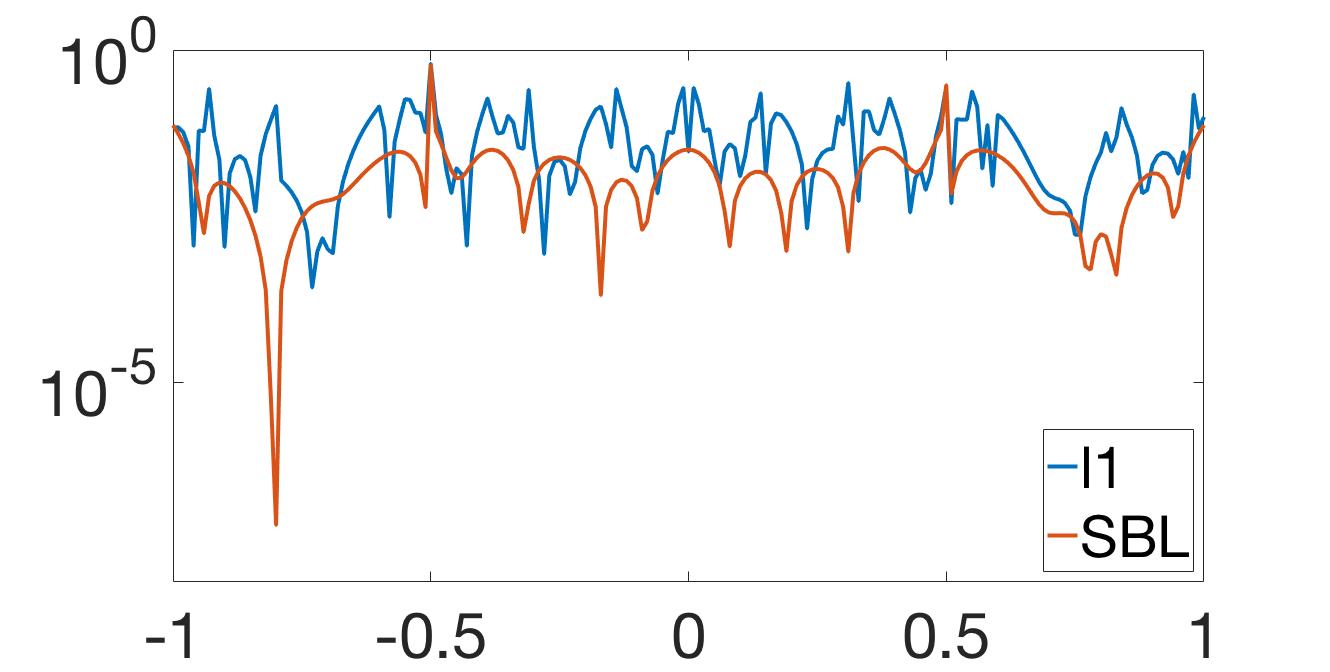}
\includegraphics[width=.32\textwidth]{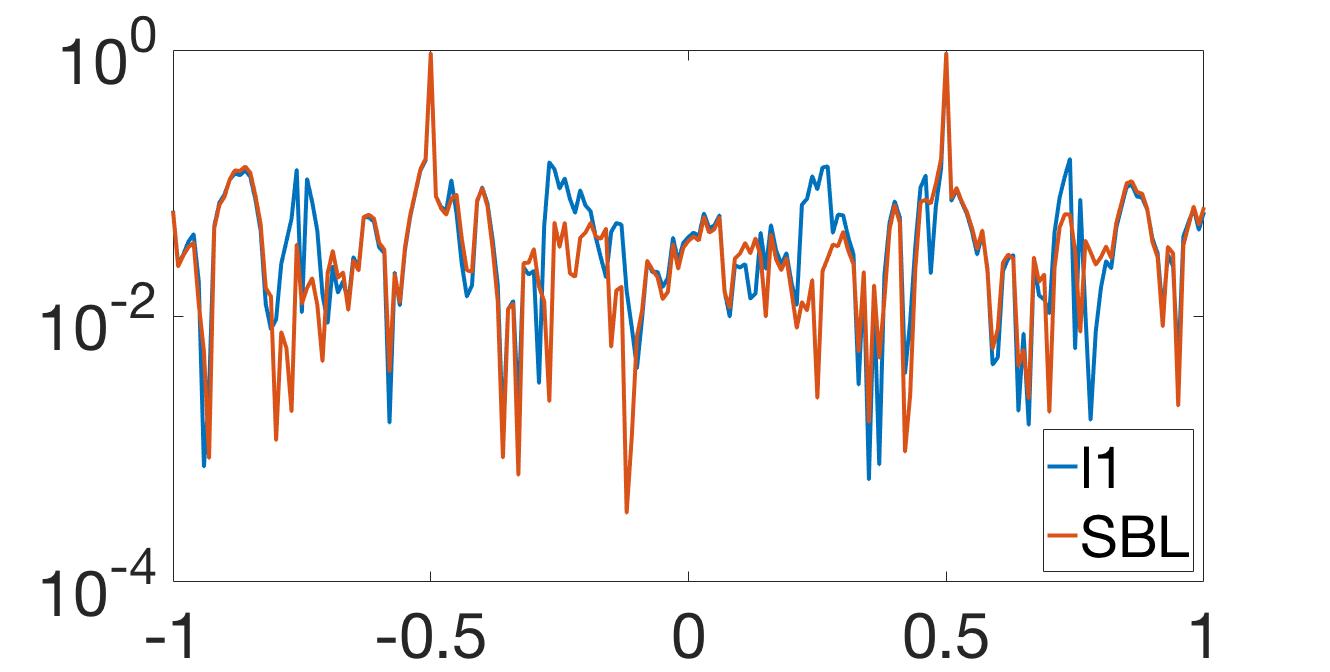}
\caption{(top row) Noisy signal reconstructions via Algorithm \ref{alg:eal2} comparing $\ell_1$ and SBL edge detection for development of the mask with (bottom row) log error for (left) $f_1$ from logarithmic sampling with $m=2$ with relative error $.1496$ and $.0823$; (center) $f_2$ from quadratic sampling and $m=3$ with relative error $.1912$ and $.1015$; and (right) $f_3$ from jittered sampling and $m=1$ with relative error $.1683$ and $.1527$.}
\label{fig:recon_noise}
\end{figure}

Since edges represent sparse features in an image, edge detection methods are often employed to aid in full piecewise-smooth signal reconstruction, \cite{archibald2016image,churchill2018edge,candes2008enhancing,chartrand2008iteratively}. For example, the methods in \cite{churchill2018edge,candes2008enhancing,chartrand2008iteratively} are designed to apply the regularization more strongly away from the edges.  As noted in the introduction, typical $\ell_1$ regularization, which can be considered as a type I MAP estimate, has been enhanced by using iterative reweighting techniques, both for $\ell_1$ and for $\ell_2$ regularization.   Such methods are particularly effective when the edges are well separated, that is, when there is sufficient resolution in the acquired data, as well as limited noise.  The performance of iterative reweighting schemes deteriorate when this is not the case, mainly because of the propagation and enhancement of false edges caused by the reweighting process. These issues inspired the development of the edge adaptive $\ell_2$ regularization method in \cite{churchill2018edge}, which is particularly appealing because it overcomes the magnitude dependence of $\ell_1$ norm regularization by directly identifying edges in one initial step. The procedure begins by creating a binary weighting matrix, or mask, based on the edge locations found using an (any) edge detection algorithm.  A weighted $\ell_2$-regularized optimization problem is then solved to achieve the full signal reconstruction.  Since the vector inside the regularization term should now be identically zero instead of sparse, it is appropriate and computationally advantageous to regularize using the $\ell_2$ norm. The effect is similar to that of skipping to the final iteration of an iteratively reweighted $\ell_2$ scheme, \cite{chartrand2008iteratively}. The full procedure is described in Algorithm \ref{alg:eal2}.  In \cite{churchill2018edge}, this method was demonstrated to have potential use in SAR image reconstruction. Here we simply demonstrate its compatibility and effectiveness with this SBL edge detection method starting from non-uniform Fourier data on piecewise smooth functions $f_1$, $f_2$, and $f_3$ given in Examples \ref{ex:ex1}, \ref{ex:ex2}, and \ref{ex:ex3} respectively. Figure \ref{fig:recon_noise} shows full signal reconstructions of these examples using a variety of sampling patterns when zero-mean Gaussian noise with standard deviation $0.02$ is added to the given Fourier data.  We use the polynomial annihilation method, \cite{archibald2005polynomial,archibald2016image}, of various orders to ensure sparsity of the corresponding jump function $[f]$. Other operators, e.g. wavelets, may also be used.

Note that in Algorithm \ref{alg:eal2} we introduce the parameter $\tau$, but in the experiments shown simply set $\tau=\frac{1}{2J+1}$, which is consistent with the grid resolution. Figure \ref{fig:threshold} also shows that our method performs well for a broad range of $\tau$.

\begin{algorithm}[h!]
\caption{Edge-adaptive $\ell_2$-regularized image reconstruction}
\label{alg:eal2}
\begin{algorithmic}[1]
\STATE Reconstruct the jump function, $[f]$, using Algorithm \ref{alg:edgel1} or \ref{alg:sbl} as $\mathbf{g}^*$.

\STATE For each index $j=-J,\ldots,J$ such that $|\mathbf{L}^m\mathbf{g}^*_{j}|>\tau$, set $\mathbf{y}_{j}=0$. Else, $\mathbf{y}_{j}=1$. Here $\tau$ is a user-defined thresholding parameter. The mask is $\mathbf{M} = \text{diag}(\mathbf{y})$, and $\mathbf{L}^m$ is the $m$th order polynomial annihilation operator, \cite{archibald2005polynomial}, similar to a high order total variation operator. It is necessary to apply $\mathbf{L}^m$ so that the stencil of the sparsity operator is accounted for in the mask.

\STATE The edge-adaptive $\ell_2$ regularization image reconstruction is the solution to the optimization problem,
\begin{eqnarray}
\label{eq:EAR}
 {\bf f}^* & = & \arg\min_{\bf f} \left\{||\mathbf{F}{\bf f} - {\bf \hat f} \rvert \rvert^2_2 +  \lambda \lvert \lvert  \mathbf{M}\mathbf{L}^m{\bf f} \rvert \rvert^2_2\right\},
\end{eqnarray}
where $\mathbf{F}$ is the non-uniform Fourier measurement matrix, $\mathbf{\hat{f}}$ is the vector of collected Fourier coefficients, and $\lambda>0$ is the regularization parameter. 
\end{algorithmic}
\end{algorithm}

\section{Conclusion}\label{sec:conclusion}
The SBL edge detection method given non-uniform Fourier data developed here compares favorably to typical $\ell_1$-regularized sparse signal recovery methods. This is true even when the data are explicitly considered in the recovery method, as in \cite{gelb2017detecting}. Moreover, this type-II  approach is more robust than typical type-I MAP estimates, which can be viewed as standard $\ell_1$ regularization methods, since the regularization parameters are estimated directly from the data and modified accordingly.  The lack of automation in standard $\ell_1$ regularization methods is considered highly unfavorable by most practitioners.  Our method can also be used to reconstruct a piecewise smooth signal as a pre-processing step for the edge-adaptive reconstruction, \cite{churchill2018edge}. Our numerical experiments demonstrate that both the edge detection and resulting reconstruction are robust with respect to noise and signal type, with our method consistently outperforming the one in  \cite{gelb2017detecting}. Future investigations will include further development of Algorithms \ref{alg:edgel1} and \ref{alg:sbl} for two-dimensional signals. While theoretically our method expands to multiple dimensions in a dimension-by-dimension manner, it is still cost prohibitive.  Some of the ideas discussed in Section \ref{Discussion} will be used to improve the efficiency of Algorithm \ref{alg:sbl}, which will be important for higher dimensional reconstructions and larger signals in general.

\bibliographystyle{acm}
\bibliography{refs}

\end{document}